\documentclass[prl,preprintnumbers,amsmath,amssymb,superscriptaddress,twocolumn]{revtex4}
\pdfoutput=1 
\usepackage{graphicx}
\usepackage{amsfonts}
\usepackage{url}
\usepackage{epstopdf}
\usepackage{color}
\usepackage{tikz-cd}
\usepackage{amsmath}
\usepackage{bm}
\usepackage{soul}
\usepackage{lipsum}
\usepackage[colorlinks= true, linkcolor=blue, citecolor=blue, urlcolor=blue]{hyperref}

\usepackage[all]{hypcap}

\def\N{\mathcal N}

\def\pa{\partial\Omega}

\def\R{{\mathbb R}}
\def\T{{\mathcal T}}

\def\r{\bm{r}}

\def\N{\mathcal N}

\def\pa{\partial\Omega}

\def\R{{\mathbb R}}
\def\T{{\mathcal T}}

\def\r{\bm{r}}

\def\Var{\mathrm{Var}}

\def\tauw{\T_{\rm ad}}

\begin{document}

\title{Escape of a Sticky Particle}

\author{Yuval Scher}
\email{yuvalscher@mail.tau.ac.il}
\affiliation{
School of Chemistry, Center for the Physics \& Chemistry of Living Systems, Ratner Institute for Single Molecule Chemistry,
and the Sackler Center for Computational Molecular \& Materials Science, Tel Aviv University, 6997801 Tel Aviv, Israel}

\author{Shlomi Reuveni}
\email{shlomire@tauex.tau.ac.il}
\affiliation{
School of Chemistry, Center for the Physics \& Chemistry of Living Systems, Ratner Institute for Single Molecule Chemistry,
and the Sackler Center for Computational Molecular \& Materials Science, Tel Aviv University, 6997801 Tel Aviv, Israel}

\author{Denis~S.~Grebenkov}
 \email{denis.grebenkov@polytechnique.edu}
\affiliation{
Laboratoire de Physique de la Mati\`{e}re Condens\'{e}e, \\ 
CNRS -- Ecole Polytechnique, Institut Polytechnique de Paris, 91120 Palaiseau, France}


\begin{abstract}  
Adsorption to a surface, reversible-binding, and trapping are all prevalent scenarios where particles exhibit ``stickiness". Escape and first-passage times are known to be drastically affected, but detailed understanding of this phenomenon remains illusive. To tackle this problem, we develop an analytical approach to the escape of a diffusing particle from a domain of arbitrary shape, size, and surface reactivity. This is used to elucidate the effect of stickiness on the escape time from a slab domain: revealing  how adsorption and desorption rates affect the mean and variance, and providing a novel approach to infer these rates from measurements. Moreover, as any smooth boundary is locally flat, slab results are leveraged to devise a numerically efficient scheme for simulating sticky boundaries in arbitrary domains. Generalizing our analysis to higher dimensions reveals that the mean escape time abides a general structure that is independent of the dimensionality of the problem. This letter thus offers a new starting point for analytical and numerical studies of stickiness and its role in escape, first-passage, and diffusion-controlled reactions.
\end{abstract}

\maketitle

Stickiness can drastically alter the completion time of random processes. A prominent example is the escape problem, also known as the exit problem, where one considers a particle searching for a hole or another way out of a compartment with otherwise impenetrable boundaries \cite{ward1993strong,grigoriev2002kinetics,singer2006narrow,straube2007reaction,reingruber2009gated,rupprecht2015exit,bressloff2015escape,grebenkov2017escape,grebenkov2019full,simpson2021mean,grebenkov2023encounter,guerin2023imperfect,meiser2023experiments}. The need to account for stickiness in such scenarios was already recognized for receptors diffusing in and out of the postsynaptic density while reversibly binding to scaffold proteins there \cite{holcman2006modeling,taflia2007dwell,maynard2023quantifying}. Similar issues arise when considering transport through the nuclear pore complex \cite{licata2009first,Hoogenboom2021physics}, the partially reversible trapping of receptors trafficking in dendrites \cite{bressloff2007diffusion}, and the 
reversible binding of calcium ions to sensors and buffer molecules within nerve terminals \cite{reva2021first}. Explicit account  of stickiness was also required in order to explain the extremely prolonged survival times of target proteins in a nanostructured on-chip device that was recently fabricated for selective protein separation using antibody–photoacid-modified Si nanopillars \cite{borberg2019light,borberg2021depletion}. As escape and first-passage times are conceptually equivalent \cite{redner2001guide,metzler2014first,klafter2011first}, stickiness is also expected to affect diffusion-controlled reactions, e.g., the time it takes a sticky ligand to find its receptor on the cell membrane.

Whether the origin of stickiness is physical, chemical or else, it can be described by the adsorption-desorption kinetics framework, which was developed over an entire century of intense research. However, the vast majority of theoretical works in this context were conducted on a macroscopic 1D model of a flat surface immersed in an infinite bulk of adsorbates \cite{langmuir1918adsorption,brunauer1938adsorption,ward1946time,sutherland1952kinetics,delahay1957adsorption,hansen1961diffusion,baret1968kinetics,miller1981solution,mccoy1983analytical,adamczyk1987nonequilibrium,miller1991adsorption,chang1995adsorption,liggieri1996diffusion,diamant1996kinetics,liu2009diffusion,foo2010insights,miura2015diffusion,miller2017dynamic,noskov2020adsorption}. Recently, a single-particle description of adsorption kinetics was introduced \cite{scher2022microscopic}, and a relation to reversible binding and association-recombination reactions \cite{agmon1984diffusion,agmon1988geminate,kim1999exact,agmon1989theory,prustel2012exact,grebenkov2019reversible,reva2021first,grebenkov2017first,lawley2019first,grebenkov2021reversible,grebenkov2022first} was made. Some authors have also ventured beyond the 1D semi-infinite case considering different geometries and higher dimensions   \cite{mysels1982diffusion,frisch1983diffusion,adamczyk1987adsorption, scher2022microscopic,reva2021first,prustel2012exact,kim1999exact,grebenkov2019reversible}. Yet, research and results were always focused on scenarios where all boundaries involved were assumed to be adsorbing, thus completely ignoring the problem of escape via an absorbing boundary. Moreover, in many real-life scenarios, domain boundaries comprise of a mixture of reflecting, absorbing, adsorbing, partially reactive \cite{collins1949diffusion,sano1979partially,weiss1986overview,szabo1984localized,singer2008partially,palreactive,grebenkov2020imperfect,Grebenkov20b} and/or gated \cite{szabo1982stochastically,spouge1996single,mercado2019first,godec2017first,scher2022continuous,kumar2022inference} parts, which highlights the need in adequate methods of analysis. 
\begin{figure}[t]
\begin{center}
\includegraphics[width=1\linewidth]{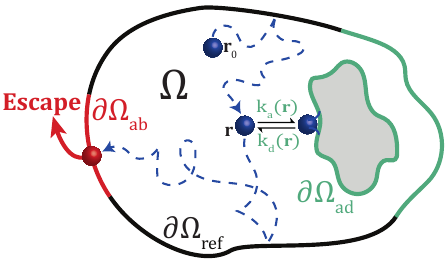}
\end{center}
\caption{
{\bf (a)} A schematic illustration of a general domain $\Omega$ with an adsorbing (sticky) surface $\partial \Omega_{\rm{ad}}$ (green), a reflecting surface $\partial \Omega_{\rm{ref}}$ (black), and an absorbing surface $\partial \Omega_{\rm{ab}}$ (red), through which the particle (blue) can escape the domain.}
\label{fig:scheme_General}
\end{figure}
%

In this Letter, we develop a general formalism to treat the escape problem of a  sticky particle diffusing in a domain of arbitrary dimension, geometry, and surface reactivity (Fig. \ref{fig:scheme_General}). We derive the partial differential equation that governs the escape time distribution for this general problem, and exemplify its solution for a particle confined to a slab with one absorbing wall and one adsorbing wall.  
In this prototypical example, we show that the mean escape time is only sensitive to the ratio of adsorption and desorption rates, while its variance is sensitive to both rates, thus allowing their inference from experimental data of escape time statistics. To gain physical intuition to the sticky escape problem, we further present a renewal approach in the slab case, which we then leverage to  construct an accurate simulation scheme for the diffusion of a
sticky particle in general domains. Finally, we solve the problems of escape from a sticky annulus and spherical shell, emphasising the universal manner in which stickiness affects the mean escape time.

\textit{The general case.}---We consider normal diffusion with diffusion coefficient $D$ in a $d$-dimensional domain $\Omega \subset \R^d$ (Fig. \ref{fig:scheme_General}), whose boundary $\pa = \partial \Omega_{\rm{ab}} \cup \partial \Omega_{\rm{ad}}\cup \partial \Omega_{\rm{ref}}$ is comprised of three disjoint parts: an absorbing part $\partial \Omega_{\rm{ab}}$ through which the particle can escape, a reflecting part $\partial \Omega_{\rm{ref}}$, and an adsorbing part $\partial \Omega_{\rm{ad}}$, allowing for reversible trapping of the particle (see below).
The propagator $p(\r,t|\r_0)$, namely the probability density to find a particle at point $\r
\in \Omega$ at time $t$ given the initial position $\r_0$, satisfies the diffusion equation:
\begin{equation} \label{eq:diffusioneq}
\partial_t p(\r,t|\r_0) = D\Delta_{\r} p(\r,t|\r_0) ,
\end{equation}
subject to the initial condition $p(\r,0|\r_0) = \delta (\r-\r_0)$, and the Dirichlet boundary conditions $p(\r,t | \r_0) = 0$ for every $\r \in \partial \Omega_{\rm{ab}}$. Here, $\Delta_{\r}$ is the Laplace operator with respect to $\r$. 
The propagator determines the probability flux density $j_{\rm{ab}}(\r,t|\r_0)=-D  \partial_n p(\r,t|\r_0)$ through a point $\r \in \partial \Omega_{\rm{ab}}$ at time $t$. This, in turn, yields the probability density function (PDF) of the escape time, $J_{\rm{ab}}(t|\r_0) = \int_{\partial \Omega_{\rm{ab}}} j_{\rm{ab}}(\r_s,t|\r_0) d \r_s$, which is also equal to the probability flux out of the compartment.

The adsorption condition on $\partial \Omega_{\rm{ad}}$ can be formulated by introducing an auxiliary probability density $\Pi (\r,t| \r_0)$ of the particle to be adsorbed to point $\r \in \partial \Omega_{\rm{ad}}$ at time $t$. We impose the following two equations for every $\r \in \partial \Omega_{\rm{ad}}$ on the adsorbing surface \cite{scher2022microscopic,grebenkov2019reversible}:
\begin{subequations}
\begin{align} \label{eq:linear_adsorption_kinetics}
 &j_{\rm{ad}}(\r,t| \r_0)  = k_a(\r) p(\r,t|\r_0) - k_d(\r) \Pi(\r,t| \r_0), \\ \label{eq:mass_balance}
&\partial_t \Pi(\r,t | \r_0)  = j_{\rm{ad}}(\r,t | \r_0)  ,
\end{align}
\end{subequations}
where $j_{\rm{ad}}(\r,t | \r_0)=-D \partial_n p(\r,t|\r_0)$. Here,  $\partial_n$ is the normal derivative oriented outwards the
domain, $k_a(\r)$ is the reactivity of the surface at the point $\r$ (characterizing the rate of adsorption from a thin reactive layer near the surface), and $k_d(\r)$ is the desorption rate at that point.  Note that the reflecting part of the boundary can be viewed as part of the adsorbing surface with zero reactivity, i.e., we can consistently define $\partial \Omega_{\rm{ref}} \subset \partial \Omega_{\rm{ad}}$, such that $k_a(\r)=0$ for every  $\r \in \partial \Omega_{\rm{ref}}$. 
Equation (\ref{eq:linear_adsorption_kinetics}) states that the diffusive flux of particles from the bulk at each point $\r \in \partial \Omega_{\rm{ad}}$ is proportional to the reactive flux on the surface, $k_a(\r) p(\r,t|\r_0)$,
minus the flux of particles that desorb from the surface.  In
turn, Eq. (\ref{eq:mass_balance}) is a mass balance equation that simply states that the uptake of adsorbed particles is given by the diffusive flux. Note that $\Pi(\r, 0| \r_0) = 0$, since motion starts in the bulk.

The Laplace transform of these two equations allows one to eliminate
$\Pi(\r,t| \r_0)$, and to reduce these equations to a single Robin-like boundary condition. Summarizing, the original problem reads in the Laplace domain as
\begin{subequations}
\begin{align} \label{eq:laplaced_diffusion_eq}
&(s - D\Delta_{\r}) \tilde{p}(\r,s|\r_0)  = \delta(\r-\r_0), \quad  \r \in \Omega , \\
\label{eq:laplaced_absorbing_BC}
 &\tilde{p}(\r,s|\r_0) = 0, \quad  \r \in \partial \Omega_{\rm{ab}},
\\ \label{eq:laplaced_adsorbing_BC}
&\partial_n \tilde{p}(\r,s|\r_0)  + q_s(\r) \tilde{p}(\r,s|\r_0)  = 0, \quad  \r \in \partial \Omega_{\rm{ad}},
\end{align}
\end{subequations}   
where tilde denotes the Laplace transform, $\tilde{p}(\r,s|\r_0) = \int\nolimits_0^{\infty} dt\,  e^{-ts} \, p(\r,t|\r_0)$, and
\begin{equation}  \label{eq:q_inhomogeneous}
q_s(\r) = \frac{k_a(\r)}{D(1 + k_d(\r)/s)}
\end{equation}
is an $s$-dependent reactivity parameter.

Conventionally the survival probability for the escape of a particle from a compartment is defined to be the spatial integral of the propagator over the entire compartment, $S_{b}(t|\r_0)=\int_{\Omega} p(\r,t|\r_0) d\r$. Here we added the subscript `$b$', standing for bulk, since survival must also include the probability to be adsorbed (hence being neither in the bulk nor escaped). The survival probability is thus given by
\begin{equation} \label{eq:survivals}
S_{\rm{ab}}(t|\r_0) := S_{b}(t|\r_0) + \Pi(t|\r_0) ,  
\end{equation}
where the subscript `$\rm{ab}$' stands for the escape through $\partial \Omega_{\rm{ab}}$, and $\Pi(t| \r_0)=\int_{\partial \Omega_{\rm{ad}}} \Pi (\r,t| \r_0)  d\r$ is the overall probability to be adsorbed at time $t$. Taking the time derivative of Eq. (\ref{eq:survivals}), one can rewrite it in terms of conservation of the probability fluxes: $J_{\rm{b}} := -\partial_t S_{\rm{b}} = J_{\rm{ab}} + J_{\rm{ad}}$, 
where we used Eq. (\ref{eq:mass_balance}). Note that this relation could also be obtained by integrating the diffusion equation (\ref{eq:diffusioneq}).
In Appendix \ref{S1} we derive the partial differential equations governing the Laplace-transformed probability fluxes $\tilde{J}_{\rm{ad}}(s|\r_0)$ and $\tilde{J}_{\rm{ab}}(s|\r_0)$. The latter is given by:
\begin{subequations}
\begin{align} \label{eq:escape_helmholtz}
&(s - D\Delta_{\r_0}) \tilde{J}_{\rm{ab}}(s|\r_0)  = 0,  \quad \r_0 \in \Omega, \\ \label{eq:escape_absorbing_bc}
&\tilde{J}_{\rm{ab}}(s|\r_0)  = 1, \hspace{2ex} \r_0 \in \partial \Omega_{\rm{ab}}, \\ \label{eq:escape_absorbing_bc2}&
\partial_{n_0} \tilde{J}_{\rm{ab}}(s|\r_0) +  q_s(\r_0) \tilde{J}_{\rm{ab}}(s|\r_0) = 0,  \quad \r_0 \in \partial \Omega_{\rm{ad}}.
\end{align}
\end{subequations}

Solving this boundary value problem for a given compartment $\Omega$ and reactivity parameter $q_s(\r)$ yields the moment-generating function of the escape time $\mathcal{T}$:
\begin{equation}
  \tilde{J}_{\rm{ab}}(s|\r_0) = \langle e^{-s\mathcal{T}}\rangle  = \int\limits_0^\infty dt e^{-ts} J_{ab}(t|\r_0). 
\end{equation}
Its derivatives give the positive integer order moments, $\langle \mathcal{T}^n\rangle = (-1)^n (\partial_s^n \tilde{J}_{\rm{ab}}(s|\r_0))_{s=0}$, while an inverse Laplace transform yields the PDF of the escape time, $J_{\rm{ab}}(t|\r_0)$. Hence, we provided a general  framework for studying the statistics of the escape time.


\begin{figure}[t]
\begin{center}
\includegraphics[width=1\linewidth]{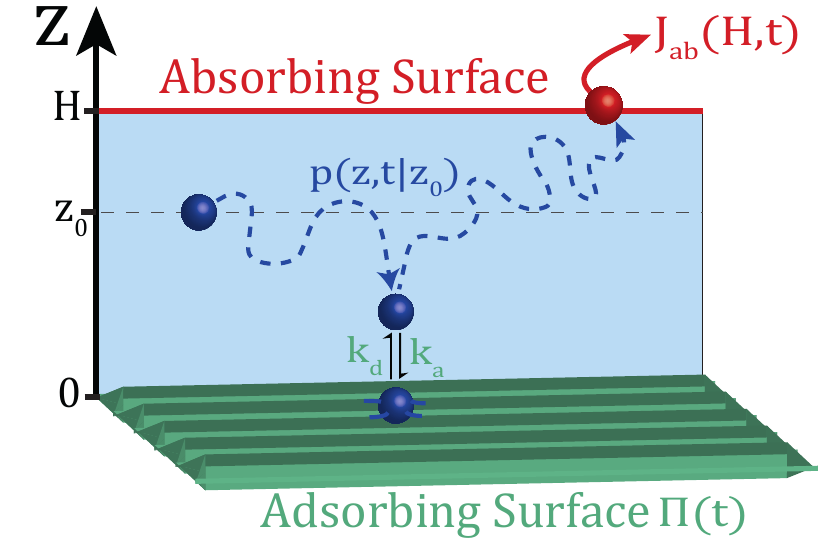}
\end{center}
\caption{ 
 A schematic illustration of a slab domain $\Omega = \R^2 \times (0,H) \subset \R^3$ with absorbing wall at $z=H$ and adsorbing wall at $z=0$. The particle starts its motion at $z=z_0$.}\label{fig:scheme_1D} 
\end{figure}  

\textit{Escape from a slab.}---We now exemplify the application of the general formalism to the case of a particle diffusing in a slab domain between two parallel planes separated by distance $H$ (Fig. \ref{fig:scheme_1D}). The boundary at $z=H$ is absorbing, while the boundary at $z=0$ is adsorbing with  $k_a$ and $k_d$. For the slab domain, Eq. (\ref{eq:q_inhomogeneous}) simplifies to $q_s = k_a/[D(1 + k_d/s)]$. This setting is equivalent to diffusion on the interval $(0,H)$ with adsorbing and absorbing endpoints at $0$ and $H$, and we are interested in getting the PDF $J_{\rm {ab}}(t|z_0)$ of the escape time from the slab, i.e., the first-passage time to $H$.

In this case, equations (\ref{eq:laplaced_diffusion_eq})-(\ref{eq:laplaced_adsorbing_BC}) for the propagator simplify as $\r=z$, $q_s(\r)=q_s$ and $\partial_n = -\partial_z$ at $z=0$. 
One divides the solution into two restrictions $z>z_0$ and $z<z_0$ and further imposes continuity of the densities and the fluxes at $z=z_0$. This yields:
\begin{align}
 \tilde{p}(z,s| z_0) =
\begin{cases}
\frac{g(z_0,s)}{g(H,s)} \frac{\sinh (\alpha  (H-z))}{D\alpha}, &  z>z_0, \\ 
  \frac{\sinh \left(\alpha  \left(H-z_0\right)\right)}{D\alpha } \frac{g(z,s)}{g(H,s)} , & z<z_0  ,
 \end{cases} 
\end{align}
where $g(x,s) := \alpha  \cosh \left(\alpha  x \right) + q_s \sinh \left(\alpha  x \right)$ and $\alpha=\sqrt{s/D}$. Since in the one-dimensional case $\tilde{J}_{\rm{ab}}(s|z_0)= - D \partial_z \tilde{p}(z,s|z_0)_{z=H}$, we have
\begin{equation} \label{eq:escapePDF_1D}   \tilde{J}_{\rm{ab}}(s|z_0) = \frac{g( z_0,s)}{g(H,s)},
\end{equation}
for the escape time probability density in the Laplace domain. Note that $\tilde{J}_{\rm{ab}}(s|z_0)$ is a solution of Eq. (\ref{eq:escape_helmholtz}) under the boundary conditions in Eqs. (\ref{eq:escape_absorbing_bc})-(\ref{eq:escape_absorbing_bc2}), and we could have bypassed the calculation of the propagator.

The escape time PDF is obtained by inverse Laplace transforming Eq. (\ref{eq:escapePDF_1D}) (see details in Appendix \ref{S2}):
\begin{align} \label{eq:1D_PDF_inversion} 
& J_{\rm{ab}}(t|z_0) = 
\sum\limits_{n=0}^\infty e^{-\beta_n^2 Dt/H^2} \times \\
&
\frac{2D \beta_n}{H^2} \biggl[\beta_n \cos(\beta_n z_0/H) + \frac{\kappa_a \beta_n^2}{\beta_n^2-\kappa_d } \sin(\beta_n z_0/H) \biggr]\times \nonumber\\
& 
\footnotesize \biggl[ \biggl(1 + 
\frac{2\kappa_a \kappa_d}{(\beta_n^2 - \kappa_d)^2}\biggr) \beta_n \sin\beta_n  -  \biggl(1 + \frac{\kappa_a \beta_n^2}{\beta_n^2 - \kappa_d}\biggr) \cos\beta_n \biggr]^{-1} , \nonumber
\end{align} \normalsize
where $\beta_n$ are the positive solutions of the transcendental equation $\beta \tan(\beta) = (\kappa_d - \beta^2)/\kappa_a$
, with $\kappa_a = k_a H /D $ and $\kappa_d = k_d H^2/D$ standing for the dimensionless adsorption and desorption rate constants. 
 \begin{figure}
\centering        
\includegraphics[width=\linewidth]{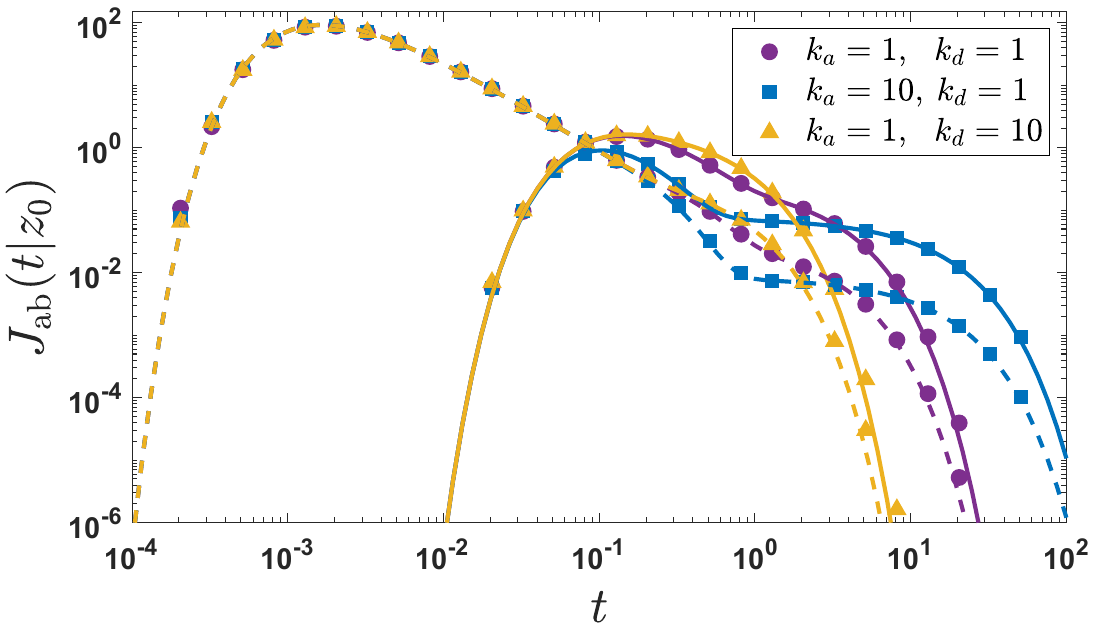}
\caption{PDF $J_{\rm{ab}}(t|z_0)$ of the escape time from the slab domain illustrated in Fig. \ref{fig:scheme_1D}. Here, we fix units of length and time by setting $H = 1$, $D = 1$, and values for $k_a$ and $k_d$ are given in the legend. Solid lines (resp., dashed lines) are the PDFs for $z_0=0.1$ (resp., $z_0=0.9$). Symbols come from Monte Carlo simulations with $10^6$ particles and a simulation time step $\Delta t=10^{-6}$ (see Appendix \ref{S4} for details).}
\label{fig:PDFfigure}
\end{figure}

 Figure \ref{fig:PDFfigure} shows the behavior of $J_{\rm{ab}}(t|z_0)$ for different values of $k_a$, $k_d$ and $z_0$. Markedly, the rates change the shape of the PDF at intermediate timescales.  
The short-time behavior is determined by ``direct trajectories'' going to the absorbing wall (the escape region). One gets the typical L\'evy-Smirnov short-time asymptotics $J_{\rm{ab}}(t|z_0) \simeq (H-z_0) e^{-(H-z_0)^2/(4Dt)}/\sqrt{4\pi Dt^3}$, that does not depend on the adsorption/desorption rates. Conversely, the long-time decay is exponential and controlled by the smallest eigenvalue $D \beta_0^2/H^2$. To see that this strongly depends on adsorption kinetics, consider the asymptotic behavior of the transcendental equation as $\beta\to 0$, from which one gets $\beta_0^2 \approx \kappa_d/(1 + \kappa_a)$. This approximation is valid whenever $\beta_0$ is small, i.e., when $\kappa_d$ is small or/and $\kappa_a$ is large. Indeed, the adsorbing wall plays the role of a temporal trap for the particle: when the particle reaches this wall, it can be easily trapped but requires long time for release. As a consequence, the  survival probability decays exponentially, with the decay time $T = H^2 /(D\beta_0^2)$, which can be very large. In Appendix \ref{S2}, the dependence of $\beta_0$ on the adsorption-desorption rates is  further discussed.

As the Laplace transform $\tilde{J}_{\rm{ab}}(s|z_0)$ is the moment generating function of the escape time, we deduce its mean
\begin{equation}     \label{eq:mean}
     \langle \mathcal{T} \rangle = \frac{H^2 - z_0^2}{2D} + \frac{ K(H-z_0)}{ D},
\end{equation}
where $K:=k_a/k_d$ is the equilibrium constant. The first term in Eq. (\ref{eq:mean}) corresponds to the mean escape time from an interval $(0,H)$ with reflecting endpoint $0$ and absorbing endpoint $H$. In turn, the second term accounts for adsorption-desorption kinetics.

A common experimental initial condition is that of uniform distribution. Averaging over the initial position in Eq. (\ref{eq:escapePDF_1D}), and calculating the first moment yields 
\begin{equation} \label{eq:mean_uniform}
    \langle \mathcal{T}_u\rangle = \frac{H^2}{3D} +  \frac{K H}{2 D},
\end{equation}
where the subscript `$u$' denotes the uniform distribution of the initial position.

\textit{Renewal Approach.}---To gain a deeper understanding of the relation in Eq. (\ref{eq:mean}), we stress that the escape time $\mathcal{T}$ is the sum of the (random) diffusion time $\mathcal{T}_d$ on the interval $(0,H)$ with reflecting endpoint $0$, and the total waiting time in the adsorbed state. Denoting the random number of adsorption events $\mathcal{N}$ and their independent, identically distributed random durations as $\mathcal{T}_w^1, \ldots, \mathcal{T}_w^{\mathcal{N}}$, we write
\begin{equation}    \label{eq:mean_RV}
     \mathcal{T} = \mathcal{T}_d + \sum_{i=1}^\mathcal{N} \mathcal{T}^i_w. 
\end{equation}
Taking mean of both sides we have $\langle\mathcal{T} \rangle =\langle \mathcal{T}_d \rangle + \langle \mathcal{N} \rangle  \langle \mathcal{T}_w \rangle$, where $\langle \mathcal{T}_w \rangle$ denotes the mean of $\mathcal{T}^i_w$.
Comparing this result with Eq. (\ref{eq:mean}), we can identify $\langle \mathcal{T}_d \rangle = (H^2 - z_0^2)/2D$ and $\langle \mathcal{T}_w \rangle = k_d^{-1}$, which gives $\langle \mathcal{N} \rangle = k_a (H - z_0)/D$.

In fact, Eq. (\ref{eq:mean_RV}) suggests that the slab case can be readily analyzed using a renewal approach, where the distribution of $\mathcal{N}$ is given in terms of the splitting probability $\mathcal{E}_{H}(z_0)$, namely the probability of escaping the compartment \textit{without any adsorption event}, given the starting position $z_0$. It is thus enough to solve the corresponding problem of diffusion in the slab, but without desorption. For example, we can generally write $ \langle \mathcal{N} \rangle  =  \left(1-\mathcal{E}_{H}(z_0)\right)/\mathcal{E}_{H}(0) $, which admits the following interpretation: Starting from $z_0$, there is a probability $1-\mathcal{E}_{H}(z_0)$ that the particle was adsorbed before escaping such that $\mathcal{N} > 0$. If so, the particle will desorb to position $z=0$, from which the probability of escaping before re-adsorbing is given by $\mathcal{E}_{H}(0)$. Essentially, adsorption at the lower boundary is a renewal moment that will repeat again and again until the particle finally escapes. We have a sequence of trails geometrically distributed with success probability $\mathcal{E}_{H}(0)$, and so the mean number of trials, each corresponding to a re-adsorption event, is simply $1/\mathcal{E}_{H}(0)$. Indeed, by plugging $\mathcal{E}_H(z_0)=(D+k_a z_0)/(D+k_a H)$, which is the splitting probability of the problem considered above, we retrieve $\langle \mathcal{N} \rangle = k_a (H - z_0)/D$.

For a more detailed discussion of the renewal approach see Appendix \ref{S3}; therein, we generalize the model by allowing for an arbitrary waiting time distribution $\psi(t)$ in the adsorbed state, and different spatial dynamics with general $\mathcal{T}_d$ and $\mathcal{E}_{H}(z_0)$. We derive generalized formulas for the PDF, the mean and the variance of the escape time, and the correlation between the diffusion time and the total waiting time. 

\begin{figure}[t]
\centering     
\includegraphics[width=1\linewidth]{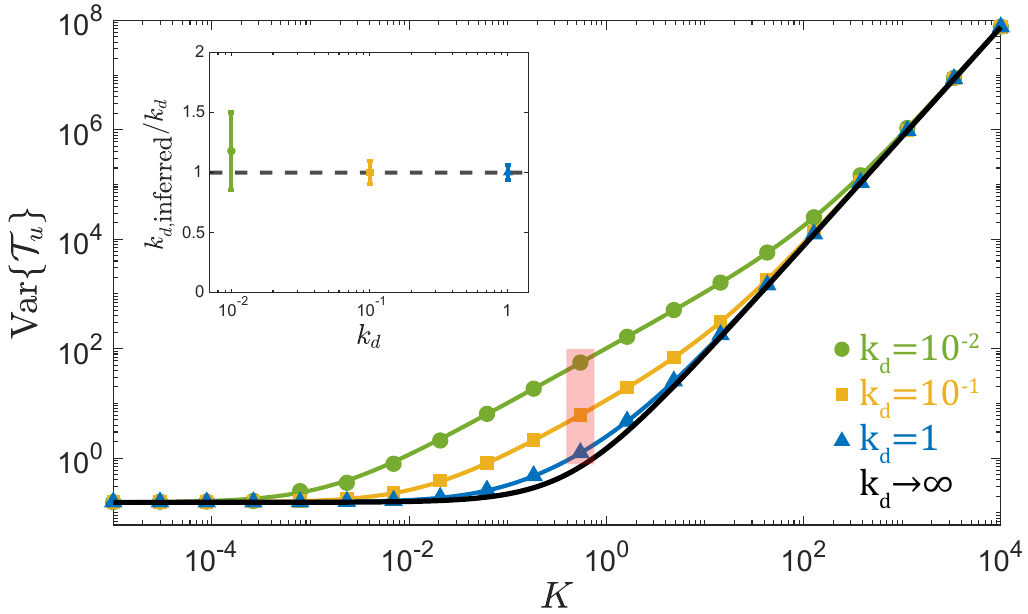}
\caption{Variance of the escape time vs. the equilibrium constant $K=k_a/k_d$ for different values of $k_d$, with $H$ and $D$ set to $1$. Here, we mimic common experimental conditions by taking the initial position distribution to be uniform. The solid lines are plotted using Eq. (\ref{eq:variance_uniform}). Symbols come from Monte Carlo simulations with $10^7$ particles. The black curve represents the fast desorption limit ($k_d \to \infty$), for which the fourth term in $\operatorname{Var}\{\mathcal{T}_u\}$ vanishes. The inset shows a successful implementation of the suggested inference scheme. To come closer to experimental conditions, we only simulated $10^4$ particles (here again $\Delta t=10^{-6}$). The relative error of the inferred $k_d$ values is plotted vs. the true $k_d$s that were fed into the simulations. All $k_d$ values share the same $K=0.43$, marked by the red strip. Error bars were calculated by repeating this procedure $10^2$ times. A detailed analysis of the statistical error can be found in Appendix \ref{S5}.}
\label{fig:Error}
\end{figure}

\textit{Inference of the adsorption-desorption rates.}---While the mean escape time in Eq. (\ref{eq:mean}) is only sensitive to the equilibrium constant $K$, we find that the variance carries information on the adsorption-desorption rates themselves:
\begin{align}\label{eq:var}
 \operatorname{Var}\{\mathcal{T}\}  =&\frac{H^4-z_0^4}{6 D^2}+\frac{2}{3} \frac{K\left(H^3-z_0^3\right)}{ D^2}  \\&+\frac{K^2\left(H^2-z_0^2\right)}{ D^2} + \underbrace{\frac{2 K (H-z_0)}{k_d D}}_{k_d \text{ sensitive}} . \nonumber
\end{align}\normalsize
Similarly,
\begin{equation}  \label{eq:variance_uniform} 
   \operatorname{Var}\{\mathcal{T}_u\} = \frac{7H^4}{45D^2} + \frac{7KH^3}{12D^2} + \frac{3H^2K^2}{4D^2} + \frac{HK}{k_d D},  
\end{equation}
where the subscript `$u$' denotes re-derivation of Eq. (\ref{eq:var}) according to a uniform distribution of the initial position.
As a consequence, one can infer $K=k_a/k_d$ by measuring the mean in Eq. (\ref{eq:mean}), and then extract $k_a$ and $k_d$ from the measured variance. In Fig. \ref{fig:Error}, we plot $\operatorname{Var}\{\mathcal{T}_u\}$ vs $K$ for three different values of $k_d$. It can be appreciated that, for a wide range of $K$ values, the variance changes considerably with different values of $k_d$, thus allowing its inference. The inset demonstrates an application of this inference scheme on simulated data. 

\textit{Simulating an adsorbing boundary.}---When thinking of an adsorbing boundary, one usually imagines a surface layer of width $\epsilon \ll 1$, from which the particle is adsorbed with rate $k_a/\epsilon$. A standard Monte Carlo simulation would thus implement such an adsorption event, wait for a random time, and then re-inject the particle into a distanced point in the bulk to resume diffusion. However, this basic scheme misses multiple adsorption events that may occur between the first desorption moment and the escape from the layer of width $\epsilon$.  Indeed, the fractal self-similar character of Brownian motion implies that a Brownian path released on a smooth boundary bounces on this boundary infinitely many times before escaping, and accurate modeling of this dynamics would require extremely small simulation time steps. If the waiting time in the adsorbed state was comparable to the simulation time step, such missed multiple adsorptions would not matter. However, in many applications, these waiting times are macroscopically large, and omission of even a single adsorption event can result in considerable errors. 

For this reason, an accurate modeling of the diffusive dynamics in the presence of an adsorbing boundary is a challenging problem. We solve it by applying our result for the escape time from an adsorbing slab where we set $H=\epsilon$. In other words, as any smooth boundary is locally flat, the escape time from a thin layer of width $\epsilon$ can be accurately approximated by the escape time from a slab of the same width. We can thus account for multiple adsorption events by drawing random times according to the distribution in Eq. (\ref{eq:1D_PDF_inversion}), with $H=\epsilon$. In Appendix \ref{S4} we exploit insight from the renewal approach to show how this can be done efficiently via a simpler algorithm that is approximate yet highly accurate. This opens the door for accurate simulations of diffusion with adsorbing boundaries in general domains and arbitrary smoothly varying adsorption-desorption rates.

\textit{Outlook.}---When coming to solve the escape problem for a sticky particle, researchers have so far resorted to simplifying assumptions \cite{holcman2006modeling,taflia2007dwell,licata2009first,borberg2019light}. Here, we tackled this problem rigorously, providing a general framework, and demonstrating its applicability using the paradigmatic problem of escape from a sticky slab domain. 

The importance of studying this fundamental example reveals itself in the ease in which we can translate its solution to general insights on the effect of stickiness. For example, dividing the mean escape time in Eq. (\ref{eq:mean}) by the mean diffusion time $\langle \mathcal{T}_d \rangle$ we observe  that
\begin{equation}     \label{eq:mean_universal}
     \frac{\langle \mathcal{T} \rangle}{\langle \mathcal{T}_d \rangle} = 1 + \frac{K}{\xi},
\end{equation}
where $\xi=(H+z_0)/2$ is an effective length scale and $K$ is the adsorption-desorption equilibrium constant. Equation (\ref{eq:mean_universal}) which describes the ratio between the mean escape time with and without stickiness generalizes to higher dimensions and other geometries.

In Appendix \ref{S6}, we solve the two-dimensional (escape from a sticky annulus) and three-dimensional (escape from a sticky spherical shell) versions of the problem illustrated in Fig. \ref{fig:scheme_1D}. In both cases we find that the mean escape time follows the form in Eq. (\ref{eq:mean_universal}), with an  effective length scale $\xi$ determined by the geometry. From an analytical perspective, this is clearly just the tip of the iceberg: detailed analysis of other cases of interest is done elsewhere \cite{InPreparation}, and we show that the second term in Eq. (\ref{eq:mean_universal}) generalizes to $\sum_{n}  K_n/\xi_n$ which accounts for the presence of multiple sticky surfaces. We conclude that this relation for the mean escape time of a sticky particle is rather general.

Our analysis revealed that adsorption and desorption rates, which may be very hard to measure directly, can instead be inferred from the mean and variance of the escape time. This opens the door for the design of experimental setups for this purpose. For example, in a spin-off on fluorescence recovery after photobleaching, one can imagine fluorescent particles in a virtual slab, where a strong laser photobleaches fluorescence above height $H$. The normalized signal from the remaining particles amounts to the survival probability and can thus be used to extract the mean and variance of the escape time. Similarly, one can think of nuclear magnetic resonance experiments, in which a surface at height $H$ causes strong relaxation that kills the transverse magnetization of the nuclei. These and other methods, e.g., single-particle tracking, can now be coupled with the results reported herein to offer new and promising ways for probing molecular interactions.

\textit{Acknowledgments.}---Shlomi Reuveni acknowledges support from the Israel Science Foundation (grant No. 394/19). This project has received funding from the European Research Council (ERC) under the European Union’s Horizon 2020 research and innovation program (Grant agreement No. 947731). Denis Grebenkov acknowledges the Alexander von Humboldt Foundation for support within a Bessel Prize award.


\setcounter{secnumdepth}{1}

\clearpage
\onecolumngrid
\appendix
\renewcommand{\theequation}{A\arabic{equation}}
\renewcommand{\thefigure}{A\arabic{figure}}
\renewcommand{\thesection}{\Alph{section}} 
\renewcommand{\bibnumfmt}[1]{[A1]}
\renewcommand{\citenumfont}[1]{A#1}
\setcounter{equation}{0}

\section{Derivation of the partial differential equation and boundary conditions governing $\tilde{J}_{\rm{ab}}(s|\textbf{r}_0)$} \label{S1}

We can write the backward diffusion equation corresponding to Eq. (\ref{eq:laplaced_diffusion_eq}), and further note that it implies a time reversal symmetry that allows us to rewrite the boundary conditions accordingly:
\begin{subequations}
\begin{align} \label{eq:laplaced_backword_diffusion_eq}
&(s - D\Delta_{\r_0}) \tilde{p}(\r,s|\r_0)  = \delta(\r-\r_0), \quad \r_0 \in \Omega, \\
\label{eq:laplaced_backword_absorbing_BC}
 &\tilde{p}(\r,s|\r_0) = 0, \quad \r_0 \in \Omega_{\rm{ab}},
\\ \label{eq:laplaced_backword_adsorbing_BC}
&\partial_{n_0} \tilde{p}(\r,s|\r_0)  + q_s(\r_0) \tilde{p}(\r,s|\r_0)  = 0, \quad \r_0 \in \Omega_{\rm{ad}},
\end{align}
\end{subequations}
where $\partial_{n_0}$ is the normal derivative with respect to the starting point $\r_0$.
Integrating over $\r \in \Omega$ we obtain  
\begin{subequations}
\begin{align} 
&(s - D\Delta_{\r_0}) \tilde{S}_{b}(s|\r_0)  = 1, \quad \r_0 \in \Omega, \\
&\tilde{S}_{b}(s|\r_0)  = 0, \quad \r_0 \in \Omega_{\rm{ab}}, \\&
\partial_{n_0} \tilde{S}_{b}(s|\r_0) +  q_s(\r_0) \tilde{S}_{b}(s|\r_0)  = 0, \quad \r_0 \in \Omega_{\rm{ad}}.
\end{align}
\end{subequations}

It is convenient to search the
solution in the form $\tilde{J}_{b}(s|\r_0) = 1 - s \tilde{S}_{b}(s|\r_0)$,
where the new function satisfies the homogeneous modified Helmholtz
equation: 
\begin{subequations}
\begin{align} \label{eq:flux_helmholtz} 
&(s - D\Delta_{\r_0}) \tilde{J}_{b}(s|\r_0)  = 0,  \quad \r_0 \in \Omega,\\
&\tilde{J}_{b}(s|\r_0)  = 1, \hspace{2ex} \r_0 \in \Omega_{\rm{ab}}, \\&
\partial_{n_0} \tilde{J}_{b}(s|\r_0) +  q_s(\r_0) \tilde{J}_{b}(s|\r_0) = q_s(\r_0),  \quad \r_0 \in \Omega_{\rm{ad}}.
\end{align}
\end{subequations}

Note that by defining $\tilde{J}_{\rm{ab}}(s|\r_0) = 1 - s \tilde{S}_{ab}(s|\r_0)$ and using the definition in Eq. (\ref{eq:survivals}) we obtain $\tilde{J}_{b} = \tilde{J}_{\rm{ab}} + s\tilde{\Pi}=\tilde{J}_{\rm{ab}}+\tilde{J}_{\rm{ad}}$. This relation simply asserts that the overall probability flux in the compartment is the sum of the probability flux to the adsorbing boundary $\Omega_{\rm{ad}}$ and the probability flux to the absorbing boundary $\Omega_{\rm{ab}}$. From the linearity of Eq. (\ref{eq:flux_helmholtz}) it is clear that $J_b$ can be written as a sum of two functions, each obeying the same system of equations, but with one of the boundary conditions replaced with an homogeneous boundary condition. It turns out that these functions are exactly $\tilde{J}_{\rm{ad}}$ and $\tilde{J}_{\rm{ab}}$: 
\begin{subequations}
\begin{align}
\label{eq:j2_helmholtz}
&(s - D\Delta_{\r_0}) \tilde{J}_{\rm{ad}}(s|\r_0)  = 0,  \quad \r_0 \in \Omega, \\
&\tilde{J}_{\rm{ad}}(s|\r_0)  = 0, \hspace{2ex} \r_0 \in \Omega_{\rm{ab}}, \\&
\partial_{n_0} \tilde{J}_{\rm{ad}}(s|\r_0) +  q_s(\r_0) \tilde{J}_{\rm{ad}}(s|\r_0)  = q_s(\r_0),  \quad \r_0 \in  \Omega_{\rm{ad}},
\end{align}
\end{subequations}
and
\begin{subequations}
\begin{align} \label{eq:escape_helmholtz_appen}
&(s - D\Delta_{\r_0}) \tilde{J}_{\rm{ab}}(s|\r_0)  = 0,  \quad \r_0 \in \Omega, \\ \label{eq:escape_absorbing_bc_appen}
&\tilde{J}_{\rm{ab}}(s|\r_0)  = 1, \hspace{2ex} \r_0 \in \Omega_{\rm{ab}}, \\ \label{eq:escape_adsorbing_bc_appen}&
\partial_{n_0} \tilde{J}_{\rm{ab}}(s|\r_0) +  q_s(\r_0) \tilde{J}_{\rm{ab}}(s|\r_0) = 0,  \quad \r_0 \in \Omega_{\rm{ad}}.
\end{align}
\end{subequations}

Let us now supply a more rigorous proof that $\tilde{J}_{\rm{ab}}(s|\r_0)$ is governed by Eqs. (\ref{eq:escape_helmholtz_appen})-(\ref{eq:escape_adsorbing_bc_appen}). From the definition of the probability density function of the escape time as the integral of the probability flux density over the absorbing part, one gets in the Laplace domain
\begin{equation} \label{eq:Jdef}
\tilde{J}_{\rm{ab}}(s|\r_0) = \int\limits_{ \Omega_{\rm{ab}}} d\r \, \underbrace{(-D\partial_n \tilde{p})}_{=\tilde{j}_{\rm{ab}}(\r,s|\r_0)} .
\end{equation}
The backward equation (\ref{eq:laplaced_backword_diffusion_eq}) implies that
\begin{equation}  \label{eq:J_s}
(s - D\Delta_{\r_0}) \tilde{J}_{\rm{ab}}(s|\r_0) = 0, \quad \r_0\in \Omega,
\end{equation}
because the arrival point $\r$ belongs to the absorbing boundary, and thus
 $\delta(\r-\r_0) = 0$.  In addition, Eq. (\ref{eq:laplaced_backword_adsorbing_BC}) immediately implies Eq. (\ref{eq:escape_adsorbing_bc_appen}). Finally, Eq. (\ref{eq:escape_absorbing_bc_appen}) simply states that any particle that starts from the escape region immediately escapes the domain so that the escape time $\mathcal{T}$ is zero, and thus $\tilde{J}_{\rm{ab}}(s|\r_0) = \langle e^{-s\mathcal{T}} \rangle = 1$.

One can also give a formal but rather technical proof.  For this purpose, one can
multiply Eq. (\ref{eq:J_s}) by $\tilde{p}(\r,s|\r_0)$, multiply
Eq. (\ref{eq:laplaced_backword_diffusion_eq}) by $\tilde{J}_{\rm{ab}}(s|\r_0)$, subtract them and
integrate over $\r_0\in\Omega$: 
\begin{align} \nonumber
\tilde{J}_{\rm{ab}}(s|\r)  =& \int\limits_\Omega d\r_0 \, \delta(\r-\r_0) \tilde{J}_{\rm{ab}}(s|\r_0)
 = \int\limits_\Omega d\r_0 \biggl[ \tilde{J}_{\rm{ab}}(s|\r_0) (s-D\Delta_{\r_0}) \tilde{p}(\r,s|\r_0) - \tilde{p}(\r,s|\r_0) (s-D\Delta_{\r_0}) \tilde{J}_{\rm{ab}}(s|\r_0)\biggr] \\
& = \int\limits_{\pa} d\r_0 \biggl[ \tilde{p}(\r,s|\r_0) D \partial_{n_0} \tilde{J}_{\rm{ab}}(s|\r_0) - \tilde{J}_{\rm{ab}}(s|\r_0) D\partial_{n_0} \tilde{p}(\r,s|\r_0)\biggr] \\
& = \int\limits_{ \Omega_{\rm{ab}}} d\r_0 \biggl[ \underbrace{\tilde{p}(\r,s|\r_0)}_{=0} D \partial_{n_0} \tilde{J}_{\rm{ab}}(s|\r_0) - \tilde{J}_{\rm{ab}}(s|\r_0) D\partial_{n_0} \tilde{p}(\r,s|\r_0)\biggr], \nonumber
\end{align}
where we used the Green's second identity and boundary conditions.  As a
consequence, we get
\begin{equation}
\tilde{J}_{\rm{ab}}(s|\r) = \int\limits_{ \Omega_{\rm{ab}}} d\r_0 \, \tilde{J}_{\rm{ab}}(s|\r_0) \underbrace{\bigl(-D\partial_{n_0} \tilde{p}(\r,s|\r_0)\bigr)}_
{=\tilde{j}_{\rm{ab}}(\r_0,s|\r)}.
\end{equation}
The time reversal symmetry of ordinary diffusion implies that the
expression in parentheses is the probability flux density
$\tilde{j}_{\rm{ab}}(\r_0,s|\r)$.  Exchanging notations $\r\leftrightarrow
\r_0$, one gets
\begin{equation}
\tilde{J}_{\rm{ab}}(s|\r_0) = \int\limits_{ \Omega_{\rm{ab}}} d\r \, \tilde{J}_{\rm{ab}}(s|\r) \tilde{j}_{\rm{ab}}(\r,s|\r_0) .
\end{equation}
Comparing this expression with Eq. (\ref{eq:Jdef}), one sees that
$\tilde{J}_{\rm{ab}}(s|\r)$ must be $1$ for all $\r\in \Omega_{\rm{ab}}$ (more
rigorously, it follows from the uniqueness of the solution of the
modified Helmholtz equation).

\renewcommand{\theequation}{B\arabic{equation}}
\renewcommand{\thefigure}{B}
\renewcommand{\theHfigure}{B}
\renewcommand{\bibnumfmt}[1]{[B1]}
\renewcommand{\citenumfont}[1]{B#1}
\setcounter{equation}{0}
\setcounter{figure}{0}    

\section{Derivation of Eq. (\ref{eq:1D_PDF_inversion}) in the main text}\label{S2}

We now proceed to invert the Laplace transform of the escape time PDF in Eq. (\ref{eq:escapePDF_1D}). The poles of this function are determined by zeros of the denominator. Setting $\beta = i\alpha H$, such that $s = -\beta^2 D/H^2$, we get the following equation for $\beta$ (and thus for the poles):
\begin{equation}  \label{eq:beta}
    \beta \tan(\beta) = \frac{\kappa_d - \beta^2}{\kappa_a} \,,
\end{equation}
where we introduced dimensionless adsorption and desorption constants 
$\kappa_a = k_a H/D $ and $\kappa_d = k_d H^2/D$.

The left-hand side is a piecewise monotonously increasing function on the intervals $(0,\pi/2)$, $(\pi/2,3\pi/2)$, etc, whereas the right-hand side is monotonously decreasing. There is thus an infinite set of solutions of this equation that we denote as $\beta_n$. The first solution $\beta_0$ lies on the interval $(0,\pi/2)$, while each other  $\beta_n$ lies on $(\pi/2 +\pi(n-1),\pi/2 + \pi n)$. In the limit $k_a = 0$ (no adsorption), the first solution is $\beta_0 = \pi/2$, while the other solutions are $\beta_n = \pi/2 + \pi n$; they correspond to an interval $(0,H)$ with reflecting endpoint $0$ and absorbing endpoint $H$. Clearly, the adsorption mechanism yields $\beta_0$ lying between $0$ and $\pi/2$. One sees that both the adsorption and desorption rates affect $\beta_0$ and thus the decay rate in a non-trivial way. In turn, in the limit $k_d =0$ (no desorption) and $k_a \to \infty$, the first solution $\beta_0 = 0$ should be excluded, while the others have a simple form $\beta_n = \pi n$; they correspond to an interval $(0,H)$ with two absorbing endpoints. If we keep $k_d =0$ but now set a finite $0<k_a<\infty$, the solution corresponds to the case of one partially absorbing endpoint (Robin boundary condition) and one absorbing endpoint.

\begin{figure}[t]
    \centering \includegraphics[width=0.5\linewidth]{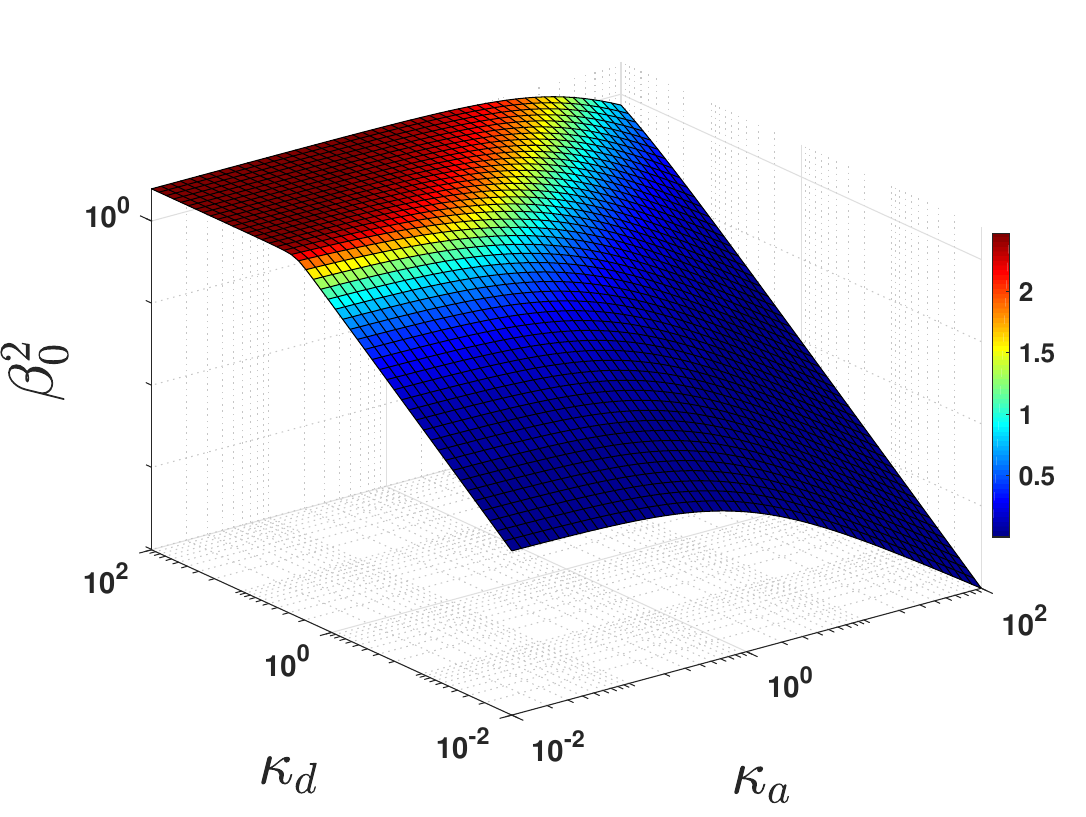}
    \caption{The first solution $\beta_0$ of Eq. (\ref{eq:beta}) that determines the smallest eigenvalue $\beta_0^2/H^2$ of the Laplace operator on the interval $(0,H)$.}
\label{fig:beta0}
\end{figure}

Figure \ref{fig:beta0} illustrates the behavior of $\beta_0^2$ as a function of the rescaled adsorption and desorption rates $\kappa_a $ and $\kappa_d$. When $\kappa_a \to 0$ and/or $\kappa_d \to \infty$, the left endpoint becomes reflecting, such that $\beta_0^2 \approx \pi^2/4$, as discussed earlier.  In the limit  $\kappa_d\to 0$, the left endpoint becomes partially reactive, with the reactivity given by $k_a$. In this case, $\beta_0^2\to 0$, but $0$ is not the pole of $\tilde{J}_{\rm ab}(s|z_0)$, so that the limiting value should be excluded, and the smallest eigenvalue is actually given by $\beta_1^2$, which is equal to $\pi^2$ in the limit $\kappa_a\to\infty$, as expected. 

To get the inverse Laplace transform of $\tilde{J}_{\rm{ab}}(s|z_0)$ we proceed to compute its residues at the poles. For this purpose, we first evaluate  
\begin{align}
 &g'(H,s)=\frac{\partial g(H,s)}{\partial s}  = \frac{\partial g}{\partial\alpha} \, \frac{d\alpha}{ds} = 
 \frac{H}{2iD\beta} \biggl[\cos\beta \biggl(1 + \frac{\kappa_a \beta^2}{\beta^2 - \kappa_d}\biggr)  - \beta \sin\beta \biggl(1 + 
\frac{2\kappa_a \kappa_d}{(\beta^2 - \kappa_d)^2}\biggr)\biggr].
\end{align}
Then, the residues are given by $ g(z_0,s_n)/g'(H,s_n)$, where $s_n = -\beta^2_n D / H^2$. Hence, applying the residue theorem to the Bromwich integral representation of the inverse Laplace transform, we get 
\begin{align} \nonumber
J_{\rm{ab}}(t|z_0) & = \frac{1}{2\pi i}\int\limits_\gamma e^{st} \tilde{J}_{\rm{ab}}(s|z_0) ds = \sum\limits_n e^{s_nt} \textrm{Res}_{s_n} \{\tilde{J}_{\rm{ab}}(s|z_0)\} \\ 
& = \sum\limits_{n=0}^\infty e^{-\beta_n^2 Dt/H^2} \frac{i}{H g'(H,s_n)} \biggl(\beta_n \cos(\beta_n z_0/H)  + \frac{\kappa_a \beta_n^2}{\beta_n^2-\kappa_d } \sin(\beta_n z_0/H) \biggr), 
\end{align}
which is equivalent to Eq. (\ref{eq:1D_PDF_inversion}).

\renewcommand{\theequation}{C\arabic{equation}}
\renewcommand{\thefigure}{C\arabic{figure}}
\renewcommand{\bibnumfmt}[1]{[C1]}
\renewcommand{\citenumfont}[1]{C#1}
\setcounter{equation}{0}

\section{Renewal approach}
\label{S3}

For an interval, one can implement a renewal approach.  In fact,
the PDF of the escape time can be written as 
\begin{align} \label{eq:Jab}
J_{\rm{ab}}(t|z_0) & = j_H(t|z_0) + \int\limits_0^t dt_1 j_0(t_1|z_0) \int\limits_{t_1}^t dt'_1 \psi(t'_1-t_1) j_H(t-t'_1|0) \\
& + \int\limits_0^t dt_1 j_0(t_1|z_0) \int\limits_{t_1}^t dt'_1 \psi(t'_1-t_1) \int\limits_{t'_1}^t dt_2 j_0(t_2-t'_1|0)  \int\limits_{t_2}^t dt'_2 \psi(t'_2-t_2) j_H(t-t'_2|0) + \ldots \nonumber
\end{align}
where $j_H(t|z_0)$ and $j_0(t|z_0)$ are the probability fluxes from the bulk
at $z = H$ and $z = 0$, and $\psi(t)$ is the
probability density of the waiting time in the adsorbed state. We emphasize that these are the fluxes from the bulk, i.e., $j_0(t|z_0)$ here does not contain the contribution from desorption (unlike $j_{\rm{ad}}(t|z_0)$ defined in the main text) -- this contribution is taken care of by $\psi(t)$. Thus, the fluxes from the bulk can be equivalently defined as the fluxes out of an interval $(0,H)$ with absorbing endpoint at $H$ and partially
reactive ($k_a>0, k_d=0$) endpoint at $0$. The
first term in Eq. (\ref{eq:Jab}) represents trajectories that escaped the domain without any
adsorption, the second term accounts for a single adsorption, the third term
for two adsorptions, and so on.  In Laplace domain, one gets
\begin{align}  
\tilde{J}_{\rm{ab}}(s|z_0)  = \tilde{j}_H(s|z_0) + \tilde{j}_0(s|z_0) \tilde{\psi}(s) \tilde{j}_H(s|0) + \ldots   \label{eq:Htilde_renewal} = \tilde{j}_H(s|z_0) + \frac{\tilde{j}_0(s|z_0) \tilde{\psi}(s) \tilde{j}_H(s|0)}{1 - \tilde{j}_H(s|0) \tilde{\psi}(s)} \,,
\end{align}
where we summed the geometric series.

For the interval $(0,H)$ with an absorbing endpoint at $H$ and a partially
reactive (note again, no desorption) endpoint at $0$, one has 
\begin{align}
\tilde{j}_H(s|z_0) & = \frac{\alpha \cosh(\alpha z_0) + q \sinh(\alpha z_0)}{\alpha \cosh(\alpha H) + q \sinh(\alpha H)} \,, \\
\tilde{j}_0(s|z_0) & = \frac{q \sinh(\alpha (H-z_0))}{\alpha \cosh(\alpha H) + q \sinh(\alpha H)} \,,
\end{align}
where $\alpha=\sqrt{s/D}$ and $q = k_a/D$.

Let us now consider an adsorbing boundary at $0$ with exponentially distributed waiting time, $\psi(t) = k_d
e^{-k_d t}$ so that $\tilde{\psi}(s) = 1/(1 + s/k_d)$.  Substituting
these expressions in Eq. (\ref{eq:Htilde_renewal}), we retrieve Eq. (\ref{eq:escapePDF_1D}).

\subsection{Mean and variance of the escape time}

In the limit $s\to 0$, we get 

\begin{align} \label{eq:splittingH}
\tilde{j}_H(s|z_0) & = \mathcal{E}_H(z_0)\bigl[1 - s \langle \mathcal{T}_H(z_0) \rangle + O(s^2)\bigr], \\ \label{eq:splitting0}
\tilde{j}_0(s|z_0) & = \mathcal{E}_0(z_0)\bigl[1 - s \langle \mathcal{T}_0(z_0) \rangle + O(s^2)\bigr],
\end{align}
where
\begin{equation} \label{eq:splittingprob}
\mathcal{E}_H(z_0) = \frac{1+qz_0}{1+qH}  \,, \qquad  \mathcal{E}_0(z_0) = \frac{q(H-z_0)}{1+qH},
\end{equation}
are the splitting probabilities on the endpoints $H$ and $0$ (e.g.,
$\mathcal{E}_H(z_0)$ is the probability of absorption on the endpoint $H$;
evidently, $\mathcal{E}_H(z_0) + \mathcal{E}_0(z_0) = 1$), and 
\begin{align} 
\langle \mathcal{T}_H(z_0) \rangle & = \frac{H-z_0}{6D(1+qH)(1+qz_0)}\biggl[ 3(H+z_0) + q(H+z_0)^2 
 + 2qHz_0 + q^2 Hz_0(H+z_0) \biggr] , \\
\langle \mathcal{T}_0(z_0) \rangle & = \frac{2H^2 + 2Hz_0 + 2H^2qz_0 -z_0^2 -z_0^2qH}{6D(1+qH)},
\end{align}
are the {\it conditional} mean absorption times to the endpoints
$0$ and $H$, respectively (e.g., $\langle \mathcal{T}_H(z_0) \rangle$ is the MFPT to the endpoint
$H$, which is conditioned by the arrival onto this endpoint).  Note
that the conditional form is necessary here because $j_H(t|z_0)$ is not
normalized to $1$ since the particle may be absorbed on the endpoint
$0$.  
Substituting these expansions into Eq. (\ref{eq:Htilde_renewal}), we
get 
\begin{align} \label{eq:expandingterms}
\tilde{J}_{\rm{ab}}(s|z_0) & \approx \mathcal{E}_H(z_0)(1 - \langle \mathcal{T}_H(z_0) \rangle s) + \frac{\mathcal{E}_0(z_0)(1 - \langle \mathcal{T}_0(z_0) \rangle s) \mathcal{E}_H(0)(1 - \langle \mathcal{T}_H(0)\rangle s)}{(1 + \langle \mathcal{T}_w \rangle s) - \mathcal{E}_0(0) (1-  \langle \mathcal{T}_0(0) \rangle s)} \,,
\end{align}
where we used $\tilde{\psi}(s) = 1 - s \langle \mathcal{T}_w \rangle + O(s^2)$, and $\langle \mathcal{T}_w \rangle$ is the
mean waiting time in the adsorbed state. Using $1-\mathcal{E}_0(0) = \mathcal{E}_H(0)$,
we then get $\tilde{J}_{\rm{ab}}(s|z_0) = 1 - \langle \mathcal{T}(z_0) \rangle s + O(s^2)$, with 
\begin{align}  \label{eq:Tz_general}
\langle \mathcal{T}(z_0) \rangle & = \biggl(\mathcal{E}_H(z_0) \langle \mathcal{T}_H(z_0) \rangle + \mathcal{E}_0(z_0) \langle \mathcal{T}_0(z_0) \rangle\biggr) + \mathcal{E}_0(z_0) \biggl[\langle \mathcal{T}_H(0) \rangle + \frac{\langle \mathcal{T}_w \rangle + \mathcal{E}_0(0) \langle \mathcal{T}_0(0)\rangle }{\mathcal{E}_H(0)}\biggr].
\end{align}
The first term is the (unconditional) mean escape time from an interval $(0,H)$ with absorbing endpoint at $H$ and partially
reactive ($k_a>0, k_d=0$) endpoint at $0$. This is equal to
\begin{align}\label{eq:26}
& \mathcal{E}_H(z_0) \langle \mathcal{T}_H(z_0) \rangle + \mathcal{E}_0(z_0) \langle \mathcal{T}_0(z_0) \rangle = \frac{(H+  z_0(1+q H))(H-z_0)}{2D(1+qH)} = \frac{H^2-z_0^2}{2D} - \mathcal{E}_0(z_0) \frac{H^2}{2D} \,,
\end{align}
and note that one recovers the results for a fully reflecting and fully absorbing boundaries at $z=0$ by setting $\mathcal{E}_0(z_0)=0$ and $\mathcal{E}_0(z_0) = \frac{H-z_0}{H}$, respectively.
As a consequence, we have 
\begin{align}
\langle \mathcal{T}(z_0) \rangle & = \frac{H^2-z_0^2}{2D} + \mathcal{E}_0(z_0) \biggl[\langle \mathcal{T}_H(0)\rangle - \frac{H^2}{2D}
+ \frac{\langle \mathcal{T}_w \rangle + \mathcal{E}_0(0) \langle \mathcal{T}_0(0) \rangle}{\mathcal{E}_H(0)}\biggr] \nonumber\\
& = \frac{H^2-z_0^2}{2D} + \mathcal{E}_0(z_0) \biggl[\frac{\langle \mathcal{T}_w \rangle}{\mathcal{E}_H(0)} \underbrace{+ \frac{\mathcal{E}_H(0) \langle \mathcal{T}_H(0) \rangle + \mathcal{E}_0(0) \langle \mathcal{T}_0(0) \rangle}{\mathcal{E}_H(0)} - \frac{H^2}{2D}}_{=0}\biggr] \nonumber\\
& = \frac{H^2-z_0^2}{2D} + \underbrace{\mathcal{E}_0(z_0) \frac{\langle \mathcal{T}_w \rangle}{\mathcal{E}_H(0)}}_{=q(H-z_0) \langle \mathcal{T}_w \rangle} ,
\end{align}
where in going from the first to second line we used Eq. (\ref{eq:26}).
Thus, despite the complexity of the general relation (\ref{eq:Tz_general}),
most contributions compensate each other, yielding a remarkably simple
expression:
\begin{equation}  \label{eq:Tz_final}
\langle \T\rangle := \langle \mathcal{T}(z_0) \rangle = \frac{H^2-z_0^2}{2D} + q (H-z_0) \langle \mathcal{T}_w \rangle ,
\end{equation}
where $q = k_a/D$.  The first term is the MFPT from an interval $(0,H)$ with absorbing endpoint at $H$ and reflecting endpoint at $0$. In turn, the second term incorporates all contributions from the adsorption/desorption events.  The proportionality of this term to $\langle \mathcal{T}_w \rangle$ suggests that it can be interpreted as the mean cumulative waiting time in the adsorbed state, whereas the first term is the mean cumulative diffusion time in the bulk.

To justify this interpretation, let us examine a random trajectory of a particle that started from $z_0$ and arrived onto the endpoint $H$.  As previously, one can distinguish two cases by whether the diffusing particle has or has not been adsorbed on the endpoint $0$ before the escape.  In the second case, there is no waiting time, and the only contribution comes from the diffusion time.  We therefore focus on the first case where the particle has been adsorbed (at least once) on $0$ before escaping the interval.  Between the first adsorption on $0$ and the escape from the interval through the endpoint $H$, the particle experienced multiple reflections from the endpoint $0$.  After a number of reflections, it may be re-adsorbed, spend some time on $0$, be desorbed, and so on.  However, if we cut off all the waiting periods in the adsorbed state (we treat them below), the adsorption/desorption mechanism does not affect the diffusive dynamics of the particle, as if the endpoint $0$ was {\it purely reflecting}. In other words, if $\langle \mathcal{T}_w \rangle = 0$ (or, in the Markovian setting, for $k_d =\infty$), there is no effect coming from adsorption/desorption, and one retrieves the results for a reflecting boundary.  As a consequence, the diffusion time in the free state is given by the first term in Eq. (\ref{eq:Tz_final}).  

For $\langle \mathcal{T}_w \rangle \ne 0$, the diffusion time should be complemented by the total waiting time that the particle has spent in the adsorbed state.  According to our derivation, the mean total waiting time is $\mathcal{E}_0(z_0) \langle \mathcal{T}_w \rangle/\mathcal{E}_H(0)$.  How can one interpret this relation?  First of all, if the particle has escaped without any adsorption, there is no such contribution.  This explains the presence of the splitting probability $\mathcal{E}_0(z_0)$, i.e., the probability that at least one adsorption occured before escaping. After each desorption, the particle starts from $0$ and can escape the interval with the probability $\mathcal{E}_H(0)$.  Let $\chi_i$ denote a Bernoulli random variable, which takes the value $0$ (re-adsorption at $i$-th trial, i.e., failure to escape) with probability $1-\mathcal{E}_H(0)$ and the value $1$ (successful escape) with probability $\mathcal{E}_H(0)$. As all escape trials are independent, the number of Bernoulli trials before escape has a geometric distribution, with the mean $1/\mathcal{E}_H(0)$. As the particle spends in each adsorbed state on average $\langle \mathcal{T}_w \rangle$ units of time, the total mean waiting time is $\mathcal{E}_0(z) \langle \mathcal{T}_w \rangle/\mathcal{E}_H(0)$, in agreement with the second term in Eq. (\ref{eq:Tz_final}).

Let us extend the above rational to represent the
escape time $\T$ as the sum of the (random) diffusion time
$\T_d$ on the interval $(0,H)$ with reflecting endpoint $0$, and the
(random) total waiting time $\tauw$ in the adsorbed state: $\T =
\T_d + \tauw$.  The latter can be formally defined as 
\begin{equation}
\tauw = \begin{cases}  0 , \quad \textrm{with probability~}\mathcal{E}_H(z_0) , \cr 
\tau_1 , \quad \textrm{with probability~}\mathcal{E}_0(z_0) \mathcal{E}_H(0) \psi(t) , \cr
\tau_2 , \quad \textrm{with probability~}\mathcal{E}_0(z_0) (1-\mathcal{E}_H(0)) \mathcal{E}_H(0) \psi_2(t) ,\cr
\cdots \cr
\tau_k , \quad \textrm{with probability~}\mathcal{E}_0(z_0) (1-\mathcal{E}_H(0))^{k-1} \mathcal{E}_H(0) \psi_k(t) ,\cr
\cdots  \end{cases}
\end{equation}
where $\psi_k(t) = (\psi \circ \psi \circ \ldots \psi)(t)$ is the
probability density function of $\tau_k$, which is defined as a sum  of $k$ independent
waiting times (note that $\tau_1 = \mathcal{T}_w$). This time is obtained as the $k$-order convolution of PDF
$\psi(t)$.  In other words, if ${\mathcal N}$ is the (random) number
of trials before escape, governed by the geometric law with $\mathcal{E}_H(0)$,
$\tauw$ is equal to $\tau_{\N}$ (apart from the value $0$, which
corresponds to no adsorption with probability $\mathcal{E}_H(z)$).

It is important to emphasize that the random variables $\T_d$ and ${\mathcal N}$ (and thus $\tauw$) are not independent.  In fact, one
can intuitively expect that large values of ${\mathcal N}$ would
correspond to large values of $\T_d$ (i.e., more escape trials imply
longer diffusion times).  The mean values $\langle \T_d\rangle$ and
$\langle \tauw\rangle$, whose sum yields the mean escape time, can be
computed independently, despite correlations between $\T_d$ and
$\tauw$, as we did above.  In contrast, correlations affect
higher-order moments and the whole distribution.  In particular, the
variance of the escape time has three contributions:
\begin{equation}  \label{eq:Var_tau_sum}
\Var\{\T\} = \Var\{\T_d\} + \Var\{\tauw\} + 2 \biggl(\langle \T_d \tauw\rangle - \langle \T_d\rangle \langle\tauw\rangle\biggr).
\end{equation}
The first term is well-known:
\begin{equation}  \label{eq:Var_taud}
\Var\{\T_d\} = \frac{H^4-z_0^4}{6D^2} \,.
\end{equation}
We compute the mean by direct computation
\begin{align}
\langle \tauw\rangle & = \int\limits_0^\infty dt \, t\, \mathrm{pdf}(t) = \mathcal{E}_0(z_0) \mathcal{E}_H(0) \sum\limits_{n=1}^\infty (1-\mathcal{E}_H(0))^{n-1} \underbrace{\int\limits_0^\infty dt \, t\, \psi_n(t)}_{= n \langle \mathcal{T}_w \rangle} = \langle \mathcal{T}_w \rangle \mathcal{E}_0(z_0) /\mathcal{E}_H(0).
\end{align}
To compute variance we use the law of total variance and obtain
\begin{align} \label{eq:Var_tauw}
\Var\{\tauw\}  &= \langle \mathcal{N} \rangle \Var\{ \mathcal{T}_w \}  +  \Var\{ \mathcal{N} \} \langle \mathcal{T}_w \rangle^2\\
&= q\left(H-z_0\right) \Var\{ \mathcal{T}_w \}  
+ q\left(H-z_0\right)\left[1+q (H + z_0) \right]                \langle \mathcal{T}_w \rangle^2  \nonumber \\
& =  q(H-z_0) \langle \mathcal{T}_w^2 \rangle  + q^2(H^2-z^2_0) \langle \mathcal{T}_w \rangle^2,  \nonumber
\end{align}
where in moving to the second line we have plugged in the first two moments of $\mathcal{N}$, which are derived in detail in Appendix \ref{S4A} (see Eq. (\ref{Nmoments})).

Comparing Eqs. (\ref{eq:Var_tau_sum}, \ref{eq:Var_taud},
\ref{eq:Var_tauw}) with the variance of $\T$, that we obtain directly
from the small-$s$ expansion of $\tilde{J}_{\rm ab}(s|z_0)$, 
\begin{align} \label{eq:Var_T} 
\mathrm{Var}\{\T\}  = \frac{H^4-z_0^4}{6D^2}    
 + \frac{2q (H^3-z_0^3)}{3D}\langle \mathcal{T}_w \rangle + 
q^2 (H^2-z_0^2) \langle \mathcal{T}_w \rangle^2 + q (H-z_0)  \langle \mathcal{T}_w^2 \rangle \,,
\end{align}
we conclude that 
\begin{equation}
\langle \T_d \tauw\rangle - \langle \T_d\rangle \langle\tauw\rangle
= \frac{2q (H^3-z_0^3)}{3D}\langle \mathcal{T}_w \rangle .
\end{equation}
In this way, we managed to characterize correlations between the
diffusion time $\T_d$ and the total waiting time $\tauw$.

While the mean escape time in Eq. (\ref{eq:Tz_final}) depends only on $q\langle \mathcal{T}_w \rangle$,
the variance of the escape time in Eq. (\ref{eq:Var_T}) depends on both $q\langle \mathcal{T}_w \rangle$
and $q\langle \mathcal{T}_w^2 \rangle$.  For the Markovian case, the distribution
of the adsorption time is exponential and one has $\langle \mathcal{T}_w \rangle = 1/k_d$ and $\langle \mathcal{T}_w^2 \rangle = 2/k_d^2$,
that implies separate dependence on $k_a$ and $k_d$ in the variance:
\begin{align} \label{eq:Var_T2}
\mathrm{Var}\{\T\}  = \frac{H^4-z_0^4}{6D^2} + \frac{K^2 (H^2-z_0^2)}{D^2}  + \frac{2K (H^3-z_0^3)}{3 D^2} + \frac{2K(H-z_0)}{k_d D}  \,,
\end{align}
where $K = k_a/k_d$.

\subsection{Splitting probabilities in terms of the Laplace transforms of the fluxes from the bulk}
\label{sec:splitting}

The splitting probability $\mathcal{E}_H(z_0)$ is the probability that a diffusing particle, initially at $z_0$, is absorbed at $H$, without adsorbing to the surface at $0$ beforehand. The complementary splitting probability $\mathcal{E}_0(z_0)$ is the probability that the particle is adsorbed at $0$ before it manages to escape. Note that $\mathcal{E}_0(z_0)$ accounts for at least one adsorption. Thus, for the purpose of the calculation of the splitting probabilities, the adsorbing boundary can be safely replaced with a partially absorbing boundary with reactivity $k_a$ (and no desorption). In other words, one only needs the fluxes from the bulk to get $\mathcal{E}_H(z_0) = \int_0^\infty J_{H}(t|z_0) dt = \tilde{J}_{H}(s = 0|z_0)$ and $\mathcal{E}_0(z_0) = \int_0^\infty J_{0}(t|z_0) dt = \tilde{J}_{0}(s = 0|z_0)$. Note that this result was already used in Eqs. (\ref{eq:splittingH})-(\ref{eq:splitting0}). 
We thus see that all the previously derived expressions that contained splitting probabilities can be written in terms of the Laplace transform $\tilde{J}_{0}(s = 0|z_0)$ and/or $\tilde{J}_{H}(s = 0|z_0)$. In fact, the renewal approach allows one to express the solution for the problem in a compartment with an adsorbing boundary in terms of the solution for the simpler problem where the adsorbing boundaries are replaced with partially reactive boundaries of reactivity $k_a$ and a waiting time distribution, $\psi(t)$, in the adsorbed state.

\renewcommand{\theequation}{D\arabic{equation}}
\renewcommand{\thefigure}{D}
\renewcommand{\theHfigure}{D}
\renewcommand{\bibnumfmt}[1]{[D1]}
\renewcommand{\citenumfont}[1]{D#1}
\setcounter{equation}{0}
\setcounter{figure}{0}   

\section{Simulating an adsorbing boundary condition}\label{S4}
 
 \subsection{Explanation} \label{S4A}
Recall the general adsorbing boundary condition given in Eqs. (\ref{eq:linear_adsorption_kinetics})-(\ref{eq:mass_balance}). For the one-dimensional case, we have 
\begin{subequations}
\begin{align} \label{eq:linear_adsorption_kinetics_appen}
 &j_{\rm{ad}}(0,t| z_0)  = k_a p(0,t|z_0) - k_d \Pi(t| z_0), \\ \label{eq:mass_balance_appen}
&\partial_t \Pi(t | z_0)  = j_{\rm{ad}}(0,t | z_0)  .
\end{align}
\end{subequations}
We show here how a numerical simulation of such a boundary condition can be performed efficiently using insights from the renewal technique presented in Appendix \ref{S3}. 

As discussed in the last subsection of Appendix \ref{sec:splitting}, the splitting probability $\mathcal{E}_H(z_0)$ is the probability that a diffusing particle, initially at $z_0$, will be absorbed at $H$, without adsorbing to the surface at $0$ beforehand. The complementary splitting probability $\mathcal{E}_0(z_0)$ is the probability that the particle is adsorbed at least once before it manages to escape. Let $\mathcal{N}$ denote the number of adsorption events before escape. We thus have:
\begin{equation}
\mathbb{P}\{\mathcal{N}=0 \mid z_0\}=\mathcal{E}_H(z_0)=\frac{1+q z_0}{1+q H},
\end{equation}
and
\begin{equation}
\mathbb{P}\{\mathcal{N}>0 \mid z_0 \}=1-\mathbb{P}\{\mathcal{N}=0 \mid z_0 \}=\mathcal{E}_0(z_0)=\frac{q(H-z_0)}{1+qH}.
\end{equation}
Markedly, when starting from the adsorbing surface we have
\begin{equation}
\mathbb{P}\{\mathcal{N}=0 \mid z_0=0\}=\mathcal{E}_H(0)=\frac{1}{1+q H}.
\end{equation}

We now aim to compute $\mathbb{P}\{\mathcal{N}=n \mid z_0 \}$, namely the probability of $n>0$ adsorptions events prior to the escape, given the initial position $z_0$. First, the particle has to be adsorbed once, the probability of which is given by $\mathcal{E}_0(z_0)$. Immediately after that adsorption event, the particle starts diffusing again, from $z_0 = 0$, and has the probability $\mathbb{P}\{\mathcal{N}=0 \mid z_0=0\}$ to escape without adsorbing for the second time. With the complementary probability the particle will adsorb again before escape and from this point onward the process is renewed. We thus have a geometrically distributed process with “success" probability $\mathbb{P}\{\mathcal{N}=0 \mid z_0=0\}$. Overall we obtain 
\begin{equation}
\mathbb{P}\{\mathcal{N}=n \mid z_0=0\}=\mathcal{E}_0(z_0)\mathcal{E}_0(0)^{n-1}\mathcal{E}_H(0).
\end{equation}
In particular we have
\begin{equation} \label{Nmoments}
\langle\mathcal{N}\rangle=q\left(H-z_0\right), \quad\left\langle\mathcal{N}^2\right\rangle=q\left(H-z_0\right)(1+2 q H) .
\end{equation}

Let us now describe the simulation procedure of an adsorbing boundary located at $z=0$. We introduce a thin boundary layer of width $\epsilon$ near the endpoint $0$. The width  should be larger than a typical one-step displacement $\sigma$ (say, $\epsilon=5\sigma$). When the particle position $z$ is at a distance smaller than $\epsilon/2$ from the boundary, we consider that they start to “interact”. As a result of this
interaction, the particle may be adsorbed a number of times, before it escapes the layer of width $\epsilon$ (we thus consider the escape problem with $H=\epsilon$). 

The escape time distribution is given exactly in Eq. (\ref{eq:1D_PDF_inversion}), but drawing random times from this bulky expression can prove numerically taxing. Instead, we suggest an alternative algorithm that reproduces the results with excellent precision and can be easily generalized. This alternative algorithm is based on the realization that in the limit $\epsilon \ll 1$ fluctuations in the escape time from the boundary layer are mostly due to fluctuations in the time spent in the adsorbed state. Thus, it is enough to retain only the effect of fluctuations in the waiting times $\mathcal{T}_w^1, \ldots, \mathcal{T}_w^{\mathcal{N}}$. Indeed, by setting $H=\epsilon$ in Eq. (\ref{eq:Var_T}) and taking this limit we are left only with the last two terms, which are equal to $\Var\{\tauw\}$.

First, one performs a Bernoulli trial to decide whether $\mathcal{N} = 0$ (with probability $\mathcal{E}_\epsilon(z)=(1 + qz)/(1 + q\epsilon)$) or $\mathcal{N} > 0$. In the former case, the particle is relocated to a new position $z = \epsilon$, while the time counter is incremented by $(\epsilon^2 - z^2)/(2D)$, i.e., the mean time needed to escape the interval $(0,\epsilon)$ with reflecting endpoint $0$ and absorbing endpoint $\epsilon$. In turn, in the latter case, we generate the random number $\mathcal{N}=n$ of adsorptions from the geometric distribution $\mathcal{E}_0(0)^{n-1}\mathcal{E}_\epsilon(0)$ for $n = 1, 2, ...$. The particle is again relocated at $z=\epsilon$, while the time counter is incremented by 
\begin{equation}\label{eq:aprox_escape}
\frac{\epsilon^2-z^2}{2 D}+\sum_{i=1}^{\mathcal{N}} \mathcal{T}^{i}_{w} ,
\end{equation}
where $\mathcal{T}_w^1, \ldots, \mathcal{T}_w^{\mathcal{N}}$ are independent waiting times generated from the exponential law with the rate $k_d$ in accordance with the first-order desorption kinetics described by the last term in Eq. (\ref{eq:linear_adsorption_kinetics}). Since a geometric sum of independent and identically distributed exponential random variables is itself exponentially distributed, one can replace the sum in Eq. (\ref{eq:aprox_escape})
by a single exponential variable with the rate $\mathcal{E}_\epsilon(0) k_d$.
Note that Eq. (\ref{eq:aprox_escape}) captures the mean escape time from the boundary layer exactly. It also captures the variance of the escape time to  second order in the layer's width $\epsilon$, with errors being $O(\epsilon^3)$. A generalization for non-exponential waiting times is trivial: one has to generate $\mathcal{T}^{i}_{w}$ according to the given probability density $\psi(t)$. 

\begin{figure}[t] 
    \centering \includegraphics[width=0.5\linewidth]{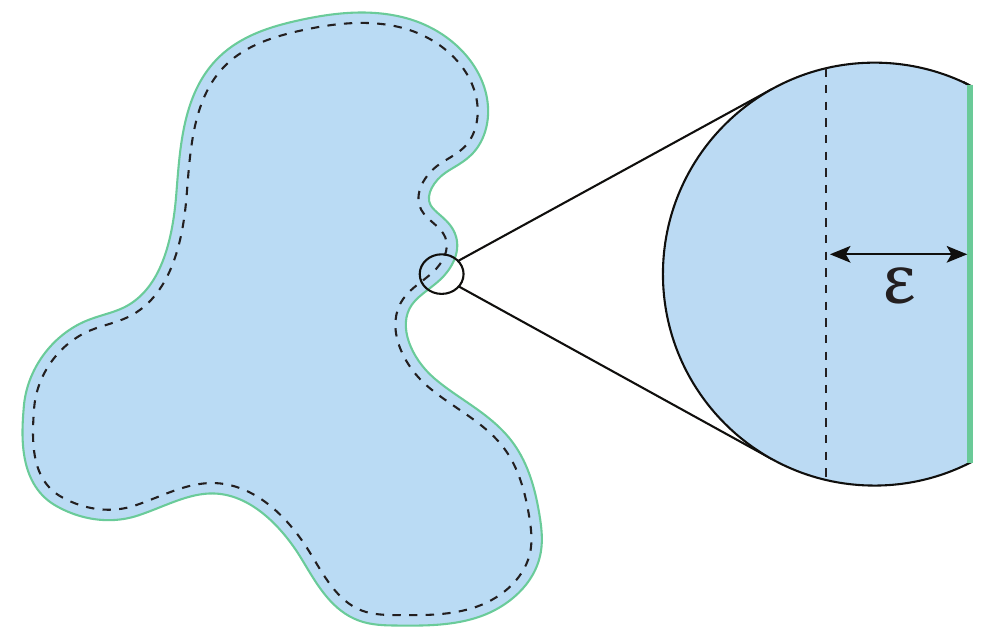}
    \caption{Simulation of an adsorbing surface of a higher-dimensional domain and of complex morphology: As any smooth boundary is locally flat, the escape time from a thin layer of width $\epsilon$ can be accurately approximated by the escape time from a slab of the same width.}\label{Fig:SimulationIllustration} 
\end{figure}

As any smooth boundary is locally flat, we can use the above procedure when simulating higher-dimensional domains (Fig. \ref{Fig:SimulationIllustration}). All we have to do is make sure that the simulation step size is small enough such that the surface can be locally approximated as a flat surface. Then, a thin subsurface layer of width $\epsilon$ can be approximated by a slab of height $\epsilon$.

\subsection{An example Matlab code}

Below we supply an example Matlab code for simulating the escape of a diffusing particle from an interval with an adsorbing boundary at $0$, and an absorbing boundary at $H$ (see Fig. \ref{fig:scheme_1D}). The simulation output is the escape time. In fact, we have used this code to simulate data shown in Fig. \ref{fig:PDFfigure} and Fig. \ref{fig:Error}. Green text denotes commentary, as we supply a short explanation of each simulation step. Note that we do not claim that the following code is of optimal performance. \\

\newpage

\noindent \textcolor{blue}{function} escape = ExampleFunction(ka, kd, H, D, z0,epsilon, dt,tmax)
\\

escape=0; \textcolor{green!50!black}{\%Define escape time.}\\

z=z0; 
\textcolor{green!50!black}{\%The location is set to the initial location.} \\

counter=0; \textcolor{green!50!black}{\%Define counter.} \\

    \textcolor{blue}{while} $1$  \textcolor{green!50!black}{\%Endless loop unless break command is performed.} \\
    
    \hspace{4ex}dz=sqrt(2*D*dt)*randn();
    z=z+dz; \textcolor{green!50!black}{\%Random step of a free diffusing particle.} \\

   \hspace{4ex} \textcolor{blue}{if} z $<$ epsilon/2  \textcolor{green!50!black}{\%When the particle position $z$ is closer to the boundary than $\epsilon/2$.}  \\
   
    \hspace{8ex} B1=rand();  \textcolor{green!50!black}{\%Generate a random number between 0 and 1.}  \\
    
    \hspace{8ex} pesc = (1 + ka*z/D)/(1 + ka*epsilon/D);  \textcolor{green!50!black}{\%The splitting probability $\mathcal{E}_\epsilon(z)$.}  \\   
    
    \hspace{8ex} B=(B1 $<$ pesc); \textcolor{green!50!black}{\%Boolean: 1 if the particle first crosses $\epsilon$ and 0 if the particle first adsorbs.}  \\ 
    
    \hspace{8ex} teps = (epsilon$\wedge$2 - z$\wedge$2)/(2*D);  \textcolor{green!50!black}{\%Time that is added to the counter later in both cases.}  \\ 
    
    \hspace{8ex} z=epsilon;  \textcolor{green!50!black}{\%The particle is relocated to $\epsilon$.}  \\

     \hspace{8ex} \textcolor{blue}{if} B==0         \textcolor{green!50!black}{\%If the particle adsorbed before crossing $\epsilon$.}  \\ 
     
     \hspace{12ex} Nads = 1 + geornd(1/(1+ka*epsilon/D),1);  \textcolor{green!50!black}{\%Number of adsorption events.}  \\ 
    
     \hspace{12ex} counter = counter + teps + sum(exprnd(1/kd,1,Nads)); \textcolor{green!50!black}{\%Update the time counter by the adsorbed} \\\vspace{1ex}
     \hspace{72ex} \textcolor{green!50!black}{\%time+time needed to diffuse a distance $\epsilon$.}

  \hspace{8ex} \textcolor{blue}{else}   \textcolor{green!50!black}{\%If crossed $\epsilon$ without first adsorbing.}  \\       
  
  \hspace{12ex} counter = counter + teps; \textcolor{green!50!black}{\%Update the time counter by the time needed to diffuse a distance $\epsilon$.}\\

\hspace{8ex} \textcolor{blue}{end}  \\   

 \hspace{4ex} \textcolor{blue}{end}  \\                

 \hspace{4ex}  \textcolor{blue}{if} counter$>$tmax  \\
 
  \hspace{8ex}          escape=tmax; \textcolor{green!50!black}{\%In case simulation power is limited. It is best to set tmax=inf.}  \\       
                            
  \hspace{8ex}          \textcolor{blue}{break};\\
                        
     \hspace{4ex}  \textcolor{blue}{elseif}   z$>$H \\
     
     \hspace{8ex}     escape=counter; \textcolor{green!50!black}{\%Save escape time and break.}  \\  
     
  \hspace{8ex}          \textcolor{blue}{break};\\
     
     \hspace{4ex} \textcolor{blue}{end}   \\

        counter=counter+dt; \textcolor{green!50!black}{\%Move time step forward.}  
        
    \textcolor{blue}{end}   

\noindent \textcolor{blue}{end}

\renewcommand{\theequation}{E\arabic{equation}}
\renewcommand{\thefigure}{E}
\renewcommand{\theHfigure}{E}
\renewcommand{\bibnumfmt}[1]{[E1]}
\renewcommand{\citenumfont}[1]{E#1}
\setcounter{equation}{0}
\setcounter{figure}{0}   

\section{Estimation of the statistical error in the inference of the desorption rate}\label{S5}

We can write our expressions for the mean and variance of the escape time $\T$ as 
\begin{align}
\label{eq:mean_appen}\mu & := \langle \T\rangle = A + BK , \\   \label{eq:variance_appen}
\sigma^2 & := \textrm{Var}\{\T\} = a + bK + cK^2 + dK/k_d,
\end{align}
where $A$, $B$, $a$, $b$, $c$, $d$ are explicitly known constants, see Eqs. (\ref{eq:mean}) and (\ref{eq:var}) for the delta-function initial condition, and the expressions given in Eqs. (\ref{eq:mean_uniform}) and (\ref{eq:variance_uniform}) for the uniform initial condition. Importantly, for both the delta-function and the uniform initial conditions the mean and variance follow the general form in Eqs. (\ref{eq:mean_appen}) and (\ref{eq:variance_appen}).

For the inference procedure, we replace the exact mean and variance by its
empirical estimates from $N$ measured escape times $\T_1, \ldots,
\T_N$: 
\begin{align}
T_1 = \frac{1}{N} \sum\limits_{n=1}^N \T_n,  \qquad
T_2 = \frac{1}{N} \sum\limits_{n=1}^N \bigl(\T_n - T_1\bigr)^2 .
\end{align}
In other words, we express:
\begin{equation}
K = \frac{T_1 - A}{B} \,, \quad k_d = \frac{dK}{T_2 - a - bK - cK^2} \,,
\end{equation}
which are now random variables due to fluctuations (statistical noise)
in the empirical estimates $T_1$ and $T_2$.  Since the fluctuations of
the empirical mean $T_1$ are characterized by standard deviation, $\Delta T_1
= \sigma/\sqrt{N}$, one has
\begin{equation}
\Delta K = \frac{\Delta T_1}{B} = \frac{\sigma}{B\sqrt{N}} \,.
\end{equation}
The fluctuations of $T_2$ are given by the standard formula from statistics,
\begin{equation}
\Delta T_2  = \sqrt{\frac{\mu_4}{N} - \frac{\sigma^4(N-3)}{N(N-1)}} \approx \frac{\sqrt{\mu_4 - \sigma^4}}{\sqrt{N}} \,,
\end{equation}
where $\mu_4 = \langle (\T - \mu)^4\rangle$ is the fourth central
moment of the escape time $\T$. 

Using the standard formulas for estimating the errors of measured
quantities, we have
\begin{equation}
\Delta k_d = \frac{d \Delta K}{T_2 - a - bK - cK^2} + \frac{dK(\Delta T_2 + b\Delta K + 2cK\Delta K)}{(T_2 - a - bK - cK^2)^2} \,,
\end{equation}
from which 
\begin{align} \nonumber
\frac{\Delta k_d}{k_d} &= \frac{\Delta K}{K} + \frac{\Delta T_2 + (b + 2cK)\Delta K}{dK} k_d  \\  \label{eq:kd_error}
&= \frac{1}{\sqrt{N}} \biggl(\frac{\sigma}{KB} + \frac{\sqrt{\mu_4 - \sigma^4} + (b + 2cK)\sigma/B}{dK} k_d\biggr).
\end{align} 
First, we note that the relative error in the estimation of the
desorption rate $k_d$ decreases as $1/\sqrt{N}$, as expected.  Let us
now inspect the coefficient in front of this factor.  For fixed $K$, in the limit
$k_d\to 0$ the first term diverges according to
Eq. (\ref{eq:variance_appen}) as $\sigma \propto 1/\sqrt{k_d}$, yielding
large relative error in the estimation of $k_d$ (note that the
absolute error $\Delta k_d$ vanishes as $\sqrt{k_d}$ in the limit
$k_d\to 0$).  Finding the behavior of the second term in
Eq. (\ref{eq:kd_error}) requires the computation of $\mu_4$ to know
its dependence on $K$ and $k_d$.  This computation is feasible but
rather tedious and is actually unnecessary.  In fact, this term can
only degrade the quality of the estimation and thus it does not alter
the above conclusion: the relative error of $k_d$ increases in the
limit $k_d\to 0$.

\begin{figure}[t!]
    \centering \includegraphics[width=0.5\linewidth]{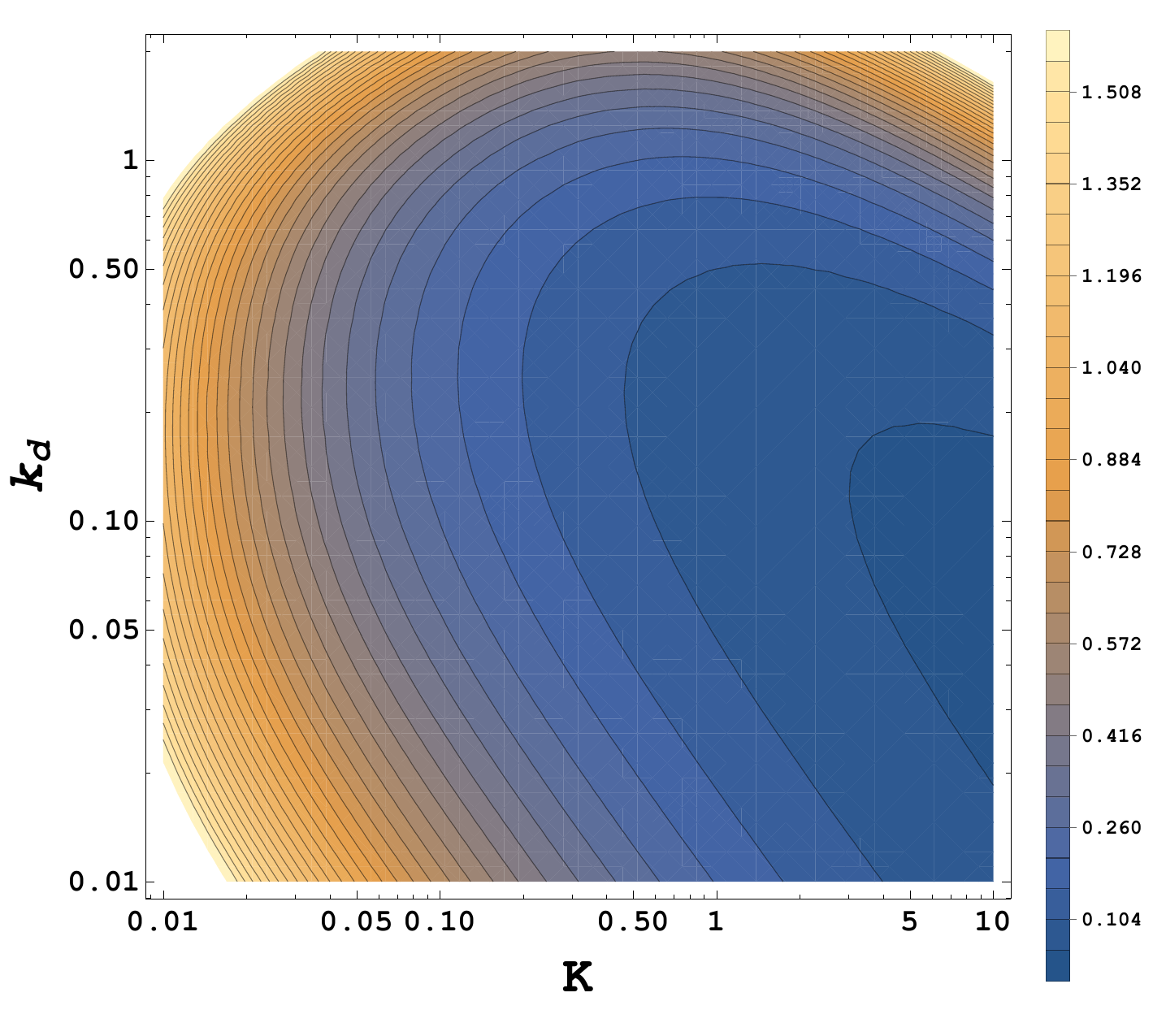}
    \caption{A contour plot of the estimated relative statistical error in the inference of $k_d$ vs. the imposed values of $K$ and $k_d$,  according to Eq. (\ref{eq:kd_error}). $H$ and $D$ were set to $1$ and a uniform initial distribution was used. This plot presents the case where the mean and variance are calculated from a sample of $10^4$ escape times. The fourth central moment $\mu_4$ of the escape time was found for the parameters at hand, using the fact that the Laplace transform in Eq. (\ref{eq:escapePDF_1D}) is the moment generating function. A general expression is computable, but it is very cumbersome and is thus not given here explicitly (it is numerically advantageous to plug the values of $D$ and $H$ in Eq. (\ref{eq:escapePDF_1D}) before computing the moments, to avoid extremely long expressions). }
\label{fig:error}
\end{figure}

Away from this limit, the relative error exhibits non-monotonous behavior. To show this we present  a contour plot of the relative error as function of $K$ and $k_d$, see Fig. \ref{fig:error}. For instance, if $K$ is fixed, the ratio $\Delta k_d/k_d$ increases in both limits $k_d\to 0$ and $k_d\to\infty$ but reaches a minimum at an intermediate value of $k_d$. Figure \ref{fig:error} can thus be used a guide for designing inference techniques, as it gives a theoretical estimation for the accuracy of the results in different regimes of the parameters space. Note that in making this figure we have assumed a uniform initial distribution and set $H$ and $D$ to 1. It can be thus seen as complementary to Fig. \ref{fig:Error}.

\renewcommand{\theequation}{F\arabic{equation}}
\renewcommand{\thefigure}{F\arabic{figure}}
\renewcommand{\bibnumfmt}[1]{[F1]}
\renewcommand{\citenumfont}[1]{F#1}
\setcounter{equation}{0}

\section{Escape of a sticky particle in  higher dimensions -- annulus and a spherical shell}\label{S6}

To give another example of an explicit computation of the probability density of the escape time, we also consider diffusion between concentric circles or spheres of radii $R_1<R_2$, such that one of them is absorbing and the other is adsorbing.  The spherical symmetry renders this problem effectively one-dimensional, and it can thus be solved in a very similar way to the escape problem from the interval, i.e., by solving the corresponding version of Eqs. (\ref{eq:escape_helmholtz})-(\ref{eq:escape_absorbing_bc2}). For instance, let us consider a three-dimensional setting and assume that the inner boundary is adsorbing and the outer is absorbing.  In this case, one still has 
\begin{align} \label{Seq:52} 
\tilde{J}_{\rm{ab}}(s|r_0) = \frac{g(r_0,s)}{g(R_2,s)},
\end{align}
with
\begin{align} \label{Seq:Laplace_3d} 
g(r,s) & = \frac{\sqrt{s/D} \cosh\left[\sqrt{s/D}(r-R_1)\right]}{r}   + \frac{(q_s + 1/R_1) \sinh\left[\sqrt{s/D}(r-R_1)\right]}{r}.
\end{align}
The poles are
determined by the condition $g(R_2,s) = 0$.  Setting $s = - D
\beta^2/(R_2-R_1)^2$, the equation $g(R_2,s) = 0$ reads 
\begin{equation*} 
0 = \frac{(R_2-R_1) R_2}{i} g(R_2,s) = \beta \cos(\beta) 
 + \biggl(\frac{\kappa_a}{1 - \kappa_d /\beta^2} + \frac{R_2-R_1}{R_1}\biggr) \sin(\beta) ,
\end{equation*}
with $\kappa_a = k_a (R_2 - R_1) /D $ and $\kappa_d = k_d (R_2 - R_1)^2/D$. This can also be written as
\begin{equation}
\frac{\beta}{\tan(\beta)} = - \biggl(\frac{\kappa_a}{1 - \kappa_d /\beta^2} + \frac{R_2-R_1}{R_1}\biggr) .
\end{equation}
The left-hand side decreases piecewise monotonously on the intervals
$(0,\pi)$, $(\pi,2\pi)$, etc., whereas the right-hand side increases
piecewise monotonously on $(0,\sqrt{\kappa_d})$ and
$(\sqrt{\kappa_d},+\infty)$.  One can therefore check that there is a
single solution on each interval $(\pi n, \pi (n+1))$ denoted as
$\beta_n$.  In turn, there are two solutions on the interval that
contains $\sqrt{\kappa_d}$, except if $\sqrt{\kappa_d} = \pi k$.
These solutions determine the poles $s_n = - D\beta_n^2/(R_2-R_1)^2$.  For
evaluating the residues, one finds
\begin{align} \nonumber
\frac{dg(R_2,s)}{ds}|_{s=s_n} & = \frac{R_2-R_1}{2DR_2 i\beta_n} \biggl\{ \biggl(1 + \frac{R_2-R_1}{R_1} + \frac{\kappa_a}{1 - \kappa_d/\beta_n^2}\biggr)
\cos\beta_n  - \biggl(1 + \frac{2\kappa_a \kappa_d}{(\kappa_d - \beta_n^2)^2}\biggr) \beta_n \sin \beta_n \biggr\} .
\end{align}
As a consequence, one gets
\begin{equation}
J(t|r) = \sum\limits_{n=0}^\infty  \frac{g\left(r,-D\beta_n^2/(R_2-R_1)^2\right)}{\frac{dg(R_2,s)}{ds}|_{s=s_n}} e^{-Dt\beta_n^2/(R_2-R_1)^2} ,
\end{equation}
where
\begin{align} \nonumber
& g\left(r,-D\beta_n^2/(R_2-R_1)^2\right) = \frac{i\beta_n \cos\left[\beta_n (r-R_1)/(R_2-R_1)\right]}{r (R_2-R_1)} \\
&  + \frac{\left[\kappa_a/(1 - \kappa_d/\beta_n^2) + (R_2-R_1)/R_1\right] i \sin\left[\beta_n (r-R_1)/(R_2-R_1)\right]}{r (R_2-R_1)} \,.
\end{align}

In two dimensions, one has
\begin{align}  \nonumber
g(r,s) & = \left[q_s K_0(R_1\sqrt{s/D} ) + \sqrt{s/D} K_1(R_1\sqrt{s/D} )\right] I_0(r \sqrt{s/D} )  \\
& -\left[q_s I_0(R_1 \sqrt{s/D} ) - \sqrt{s/D} I_1(R_1 \sqrt{s/D} )\right] K_0(r\sqrt{s/D} ) \,,
\end{align}
where and $I_0(\cdot)$ and $K_0(\cdot)$ are the modified Bessel functions of the first and second kind of order $0$. Similar computations can be performed. 

To obtain the mean escape time we recall that $\tilde{J}_{\rm{ab}}(s|r_0) = 1 - s\langle \mathcal{T} (r_0) \rangle + O(s^2)$. Taking the small-$s$ expansion of Eq. (\ref{Seq:52}), we obtain for the three-dimensional case 
\begin{equation}
\langle \mathcal{T} (r_0) \rangle = \frac{(R_2 - r_0) \left[ 6KR_1^2 - 2 R_1^3 + r_0 R_2 (r_0 + R_2) \right]}{6 D r_0 R_2}.
\end{equation}
Letting $\langle \mathcal{T}_d \rangle$ stand for the mean escape time with $K=0$, namely the case of no stickiness, we observe that
\begin{equation}\label{eq:F8}
\frac{\langle \mathcal{T} (r_0) \rangle}{\langle \mathcal{T}_d \rangle} = 1 + \frac{K}{\xi},
\end{equation}
where we identified the effective length
\begin{equation}
    \xi = \frac{r_0 R_2 (r_0 + R_2)-2R_1^3}{6 R_1^2}.
\end{equation}

\noindent Similarly, for the two-dimensional case we find
\begin{equation}
\langle \mathcal{T} (r_0) \rangle = \frac{R_2^2 - r_0^2 + 2R_1(R_1 - 2K) \ln\left( \frac{r_0}{R_2}\right)}{4 D},
\end{equation}
which also satisfies Eq. (\ref{eq:F8}), with
\begin{equation}
 \xi= \frac{r_0^2 - 2 \ln\left( \frac{r_0}{R_2}\right) R_1^2 - R_2^2}{4 R_1 \ln\left( \frac{r_0}{R_2}\right) }.
\end{equation}


\begin{thebibliography}{10}


\bibitem{ward1993strong}
M.J. Ward, J.B. Keller (1993). Strong localized perturbations of eigenvalue problems. \textit{SIAM Journal on Applied Mathematics}, 53(3), 770-798.

\bibitem{grigoriev2002kinetics}
I.V. Grigoriev, Y.A. Makhnovskii, A.M. Berezhkovskii, V.Y. Zitserman (2002). Kinetics of escape through a small hole. The \textit{Journal of chemical physics}, 116(22), 9574-9577.

\bibitem{singer2006narrow}
A. Singer, Z. Schuss, D. Holcman, R.S. Eisenberg (2006). Narrow escape, part I. \textit{Journal of Statistical Physics}, 122(3), 437-463.

\bibitem{straube2007reaction}
R. Straube, M.J. Ward, M. Falcke (2007). Reaction rate of small diffusing molecules on a cylindrical membrane. \textit{Journal of Statistical Physics}, 129, 377-405.

\bibitem{reingruber2009gated}
J. Reingruber, D. Holcman (2009). Gated narrow escape time for molecular signaling. \textit{Physical review letters}, 103(14), 148102.

\bibitem{rupprecht2015exit}
J.F. Rupprecht, O. Bénichou, D.S. Grebenkov, R. Voituriez (2015). Exit time distribution in spherically symmetric two-dimensional domains. \textit{Journal of Statistical Physics}, 158, 192-230.

\bibitem{bressloff2015escape}
P.C. Bressloff, S.D. Lawley (2015). Escape from subcellular domains with randomly switching boundaries. \textit{Multiscale Modeling \& Simulation}, 13(4), 1420-1445.

\bibitem{grebenkov2017escape}
D.S. Grebenkov, J.F. Rupprecht (2017). The escape problem for mortal walkers. The \textit{Journal of Chemical Physics}, 146(8), 084106.

\bibitem{grebenkov2019full}
D.S. Grebenkov, R. Metzler, G. Oshanin (2019). Full distribution of first exit times in the narrow escape problem. \textit{New Journal of Physics}, 21(12), 122001.

\bibitem{simpson2021mean}
M.J. Simpson, D.J. VandenHeuvel, J.M. Wilson, S.W. McCue, E.J. Carr (2021). Mean exit time for diffusion on irregular domains. \textit{New Journal of Physics}, 23(4), 043030.

\bibitem{grebenkov2023encounter}
D.S. Grebenkov (2023). Encounter-based approach to the escape problem, \textit{Physical Review E}, 107, 044105.

\bibitem{guerin2023imperfect}
T. Guérin,  M. Dolgushev, O. Bénichou, R. Voituriez (2023). Imperfect narrow escape problem. \textit{Physical Review E}, 107(3), 034134.

\bibitem{meiser2023experiments}
E. Meiser, R. Mohammadi, N. Vogel, D. Holcman, S.F. Fenz (2023). Experiments in micro-patterned model membranes support the narrow escape theory. bioRxiv, 2023-01.

\bibitem{maynard2023quantifying}
S.A. Maynard, J. Ranft, A. Triller (2023). Quantifying postsynaptic receptor dynamics: insights into synaptic function. \textit{Nature Reviews Neuroscience}, 24(1), 4-22.

\bibitem{holcman2006modeling}
D. Holcman, A. Triller (2006). Modeling synaptic dynamics driven by receptor lateral diffusion. \textit{Biophysical journal}, 91(7), 2405-2415.

\bibitem{taflia2007dwell}
A. Taflia, D. Holcman (2007). Dwell time of a Brownian molecule in a microdomain with traps and a small hole on the boundary. \textit{The Journal of chemical physics}, 126(23), 234107.

\bibitem{licata2009first}
N.A. Licata, S.W. Grill (2009). The first-passage problem for diffusion through a cylindrical pore with sticky walls. \textit{The European Physical Journal E}, 30, 439-447.

\bibitem{Hoogenboom2021physics}Hoogenboom, B.W., Hough, L.E., Lemke, E.A., Lim, R.Y., Onck, P.R. and Zilman, A., 2021. Physics of the nuclear pore complex: Theory, modeling and experiment. \textit{Physics reports}, 921, pp.1-53.

\bibitem{bressloff2007diffusion}
P.C. Bressloff, B.A. Earnshaw (2007). Diffusion-trapping model of receptor trafficking in dendrites. \textit{Physical Review E}, 75(4), 041915.

\bibitem{reva2021first}
M. Reva, D.A. DiGregorio, D.S. Grebenkov (2021). A first-passage approach to diffusion-influenced reversible binding and its insights into nanoscale signaling at the presynapse. \textit{Scientific reports}, 11(1), 1-17.

\bibitem{borberg2019light}
E. Borberg, M. Zverzhinetsky, A. Krivitsky, A. Kosloff, O. Heifler, G. Degabli, H. Peretz-Soroka, R. Satchi-Fainaro, L. Burstein, S. Reuveni, H. Diamant, V. Krivitsky and F. Patolsky (2019). Light-controlled selective collection-and-release of biomolecules by an on-chip nanostructured device. \textit{Nano letters}, 19(9), 5868-5878.

\bibitem{borberg2021depletion}
E. Borberg, S. Pashko, V. Koren, L. Burstein, F. Patolsky (2021). Depletion of highly abundant protein species from biosamples by the use of a branched silicon nanopillar on-chip platform. \textit{Analytical Chemistry}, 93(43), 14527-14536.

\bibitem{redner2001guide}
S. Redner, \textit{A guide to first-passage processes}. (Cambridge
University Press, Cambridge, England, 2001).

\bibitem{metzler2014first}
R. Metzler, S. Redner, and G. Oshanin, \textit{First-passage phenomena and their applications} (World Scientific,
Singapore, 2014), Vol. 35.

\bibitem{klafter2011first}
J. Klafter and I.M. Sokolov, \textit{First steps in random walks: from tools to applications}, (Oxford University Press,
New York, 2011).

\bibitem{langmuir1918adsorption}
I. Langmuir (1918). The adsorption of gases on plane surfaces of glass, mica and platinum. \textit{Journal of the American Chemical society}, 40(9), 1361-1403.

\bibitem{brunauer1938adsorption}
S. Brunauer, P.H. Emmett, E. Teller (1938). Adsorption of gases in multimolecular layers. \textit{Journal of the American chemical society}, 60(2), 309-319.

\bibitem{ward1946time}
A.F.H. Ward, L. Tordai (1946). Time‐dependence of boundary tensions of solutions I. The role of diffusion in time‐effects. \textit{The Journal of Chemical Physics}, 14(7), 453-461.

\bibitem{sutherland1952kinetics}
K.L. Sutherland. (1952). The kinetics of adsorption at liquid surfaces. \textit{Australian Journal of Chemistry}, 5(4), 683-696.

\bibitem{delahay1957adsorption}
P. Delahay, I. Trachtenberg (1957). Adsorption kinetics and electrode processes. \textit{Journal of the American Chemical Society}, 79(10), 2355-2362.

\bibitem{hansen1961diffusion}
R.S. Hansen (1961). Diffusion and the kinetics of adsorption of aliphatic acids and alcohols at the water-air interface. \textit{Journal of Colloid Science}, 16(6), 549-560.

\bibitem{baret1968kinetics}
J.F. Baret (1968). Kinetics of adsorption from a solution. Role of the diffusion and of the adsorption-desorption antagonism. \textit{The Journal of Physical Chemistry}, 72(8), 2755-2758.

\bibitem{miller1981solution}
R. Miller (1981). On the solution of diffusion controlled adsorption kinetics for any adsorption isotherms. \textit{Colloid and polymer science}, 259(3), 375-381.

\bibitem{mccoy1983analytical}
B.J. McCoy (1983). Analytical solutions for diffusion-controlled adsorption kinetics with non-linear adsorption isotherms. \textit{Colloid and polymer science}, 261(6), 535-539.

\bibitem{adamczyk1987nonequilibrium}
Z. Adamczyk (1987). Nonequilibrium surface tension for mixed adsorption kinetics. \textit{Journal of colloid and interface science}, 120(2), 477-485.

\bibitem{miller1991adsorption}
R. Miller, G. Kretzschmar (1991). Adsorption kinetics of surfactants at fluid interfaces. \textit{Advances in colloid and interface science}, 37(1-2), 97-121.

\bibitem{chang1995adsorption}
C.H. Chang, E.I. Franses (1995). Adsorption dynamics of surfactants at the air/water interface: a critical review of mathematical models, data, and mechanisms. \textit{Colloids and Surfaces A: Physicochemical and Engineering Aspects}, 100, 1-45.

\bibitem{liggieri1996diffusion}
L. Liggieri, F. Ravera, A. Passerone (1996). A diffusion-based approach to mixed adsorption kinetics. \textit{Colloids and surfaces A: physicochemical and engineering aspects}, 114, 351-359.

\bibitem{diamant1996kinetics}
H. Diamant, D. Andelman (1996). Kinetics of surfactant adsorption at fluid-fluid interfaces. \textit{The Journal of Physical Chemistry}, 100(32), 13732-13742.

\bibitem{liu2009diffusion}
J. Liu, P. Li, C. Li, Y. Wang (2009). Diffusion-controlled adsorption kinetics of aqueous micellar solution at air/solution interface. \textit{Colloid and Polymer Science}, 287(9), 1083-1088.

\bibitem{foo2010insights}
K.Y. Foo, B.H. Hameed (2010). Insights into the modeling of adsorption isotherm systems. \textit{Chemical engineering journal}, 156(1), 2-10.

\bibitem{miura2015diffusion}
T. Miura, K. Seki (2015). Diffusion influenced adsorption kinetics. \textit{The Journal of Physical Chemistry B}, 119(34), 10954-10961.

\bibitem{miller2017dynamic}
R. Miller, E.V. Aksenenko, V.B. Fainerman (2017). Dynamic interfacial tension of surfactant solutions. \textit{Advances in colloid and interface science}, 247, 115-129.

\bibitem{noskov2020adsorption}
B.A. Noskov, A.G. Bykov, G. Gochev, S.Y. Lin, G. Loglio, R. Miller, O.Y. Milyaeva (2020). Adsorption layer formation in dispersions of protein aggregates. \textit{Advances in colloid and interface science}, 276, 102086.

\bibitem{scher2022microscopic}
Y. Scher, O. Lauber Bonomo, A. Pal, S. Reuveni (2023). Microscopic Theory of Adsorption Kinetics. \textit{The Journal of Chemical Physics}, 158, 094107.

\bibitem{agmon1984diffusion}
N. Agmon (1984). Diffusion with back reaction. \textit{The Journal of chemical physics}, 81(6), 2811-2817.

\bibitem{agmon1988geminate}
N. Agmon, E. Pines, D. Huppert. (1988). Geminate recombination in proton‐transfer reactions. II. Comparison of diffusional and kinetic schemes. \textit{The Journal of chemical physics}, 88(9), 5631-5638.

\bibitem{agmon1989theory}
N. Agmon, G. Weiss (1989). Theory of non‐Markovian reversible dissociation reactions. \textit{The Journal of chemical physics}, 91(11), 6937-6942.

\bibitem{grebenkov2017first}
D.S. Grebenkov (2017). First passage times for multiple particles with reversible target-binding kinetics. \textit{The Journal of chemical physics}, 147(13), 134112.

\bibitem{lawley2019first}
S.D. Lawley, J.B. Madrid (2019). First passage time distribution of multiple impatient particles with reversible binding. \textit{The Journal of chemical physics}, 150(21), 214113.

\bibitem{grebenkov2021reversible}
D.S. Grebenkov, A. Kumar (2021). Reversible target-binding kinetics of multiple impatient particles. \textit{The Journal of Chemical Physics}, 156(8), 084107.

\bibitem{grebenkov2022first}
D. Grebenkov, A. Kumar (2022). First-passage times of multiple diﬀusing particles with reversible target-binding kinetics. \textit{Journal of Physics A: Mathematical and Theoretical}.

\bibitem{prustel2012exact}
T. Prüstel, M. Meier-Schellersheim (2012). Exact Green's function of the reversible diffusion-influenced reaction for an isolated pair in two dimensions. \textit{The Journal of chemical physics}, 137(5), 054104.

\bibitem{grebenkov2019reversible}
D.S. Grebenkov (2019). Reversible reactions controlled by surface diffusion on a sphere. \textit{The Journal of chemical physics}, 151(15), 154103.

\bibitem{kim1999exact}
H. Kim, K.J. Shin (1999). Exact solution of the reversible diffusion-influenced reaction for an isolated pair in three dimensions. \textit{Physical review letters}, 82(7), 1578.

\bibitem{mysels1982diffusion}
K.J. Mysels (1982). Diffusion-controlled adsorption kinetics. General solution and some applications. \textit{The Journal of Physical Chemistry}, 86(23), 4648-4651.

\bibitem{frisch1983diffusion}
H.J. Frisch, K.J. Mysels (1983). Diffusion-controlled adsorption. Concentration kinetics, ideal isotherms, and some applications. \textit{The Journal of Physical Chemistry}, 87(20), 3988-3990.

\bibitem{adamczyk1987adsorption}
Z. Adamczyk, J. Petlicki (1987). Adsorption and desorption kinetics of molecules and colloidal particles. \textit{Journal of colloid and interface science}, 118(1), 20-49.

\bibitem{collins1949diffusion}
F.C. Collins, G.E. Kimball (1949). Diffusion-controlled reaction rates. \textit{Journal of colloid science}, 4(4), 425-437.

\bibitem{sano1979partially}
H. Sano, M. Tachiya (1979). Partially diffusion‐controlled recombination. \textit{The Journal of Chemical Physics}, 71(3), 1276-1282.

\bibitem{szabo1984localized}
A. Szabo, G. Lamm, G.H. Weiss (1984). Localized partial traps in diffusion processes and random walks. \textit{Journal of statistical physics}, 34(1-2), 225-238.

\bibitem{weiss1986overview}
G.H. Weiss (1986). Overview of theoretical models for reaction rates. \textit{Journal of Statistical Physics}, 42(1), 3-36.

\bibitem{singer2008partially}
A. Singer, Z. Schuss, A. Osipov, D. Holcman (2008). Partially reflected diffusion. \textit{SIAM Journal on Applied Mathematics}, 68(3), 844-868.

\bibitem{palreactive}
A. Pal, I.P. Castillo, A. Kundu (2019).
Motion of a Brownian particle in the presence of reactive boundaries. \textit{Physical Review E}, 100, 042128.

\bibitem{grebenkov2020imperfect}
D. S. Grebenkov (2020). Imperfect diffusion-controlled reactions. In Chemical Kinetics: Beyond the Textbook (pp. 191-219).

\bibitem{Grebenkov20b}  D. S. Grebenkov (2020). Paradigm shift in diffusion-mediated surface phenomena. \textit{Phys. Rev. Lett.} 125, 078102. 

\bibitem{szabo1982stochastically}
A. Szabo, D. Shoup, S.H. Northrup J.A. McCammon (1982). Stochastically gated diffusion‐influenced reactions. \textit{The Journal of Chemical Physics}, 77(9), 4484-4493.

\bibitem{spouge1996single}
J.L Spouge, A. Szabo, G.H. Weiss (1996). Single-particle survival in gated trapping. \textit{Physical Review E}, 54(3), 2248.

\bibitem{godec2017first}
A. Godec, R. Metzler (2017). first-passage time statistics for two-channel diffusion. \textit{Journal of Physics A: Mathematical and Theoretical}, 50(8), 084001.

\bibitem{mercado2019first}
G. Mercado-Vásquez, D. Boyer (2019). First hitting times to intermittent targets. \textit{Physical Review Letters}, 123(25), 250603.

\bibitem{scher2022continuous}
Y. Scher, A. Kumar, M.S. Santhanam,  S. Reuveni (2022). Continuous gated first-passage processes. arXiv preprint arXiv:2211.09164.

\bibitem{kumar2022inference}
A. Kumar, Y. Scher, S. Reuveni, M.S. Santhanam (2022). Inference from gated first-passage times. arXiv preprint arXiv:2210.00678. 

\bibitem{InPreparation}
Y. Scher, S. Reuveni, D.S. Grebenkov (2023). Escape from textured adsorbing surfaces. \textit{In preparation}.


\end{thebibliography}
\end{document}